\documentclass[superscriptaddress,twocolumn,prb]{revtex4-1}

\usepackage{threeparttable}

\usepackage{graphicx}
\usepackage{subfigure}

\usepackage{amsmath}
\usepackage{amssymb}
\usepackage{amsthm}
\usepackage{ulem} 

\begin{document}

\title{The random phase approximation applied to ice}

\author{M. Macher}

\author{J. Klime\v{s}}

\author{C. Franchini}

\author{G.~Kresse}
\affiliation{University of Vienna, Faculty of Physics and Center for 
Computational Materials Science,
 Sensengasse 8/12, A-1090 Vienna, Austria}

\begin{abstract} 
Standard density functionals without van der Waals interactions
yield an unsatisfactory description of ice phases, specifically,
high density phases occurring under pressure are too unstable
compared to the common low density phase I$_h$ observed at ambient conditions.
Although the description is improved by using functionals that
include van der Waals interactions, 
the errors in relative volumes remain sizable.
Here we assess the random phase approximation (RPA)
for the correlation energy and compare our results to experimental
data as well as diffusion Monte Carlo data for ice. The RPA yields
a very balanced description for all considered phases, approaching
the accuracy of diffusion Monte Carlo in relative energies and
volumes. This opens a route towards a concise description
of molecular water phases on surfaces and in cavities.

\vskip 1cm
Copyright (2014) American Institute of Physics. This article may be
downloaded for personal use only. Any other use requires prior permission of the author
and the American Institute of Physics.
The following article appeared in J. Chem. Phys. {\bf 140}, 084502 (2014) and may be found at
\href{http://scitation.aip.org/content/aip/journal/jcp/140/8/10.1063/1.4865748}{http://scitation.aip.org/content/aip/journal/jcp/140/8/10.1063/1.4865748}.

\end{abstract}

\maketitle

\section{Introduction}

For the last 50 years, water and ice have always been at the forefront of research.
With the emergence and establishment of density functional theory (DFT),
there has been a slow but continuous shift from empirical force field based methods towards an
{\it ab initio} description of
ice\cite{car_ice_92,klein_05,slater_jacs_06,feibelman_08,hermann_prl_08,hamada_10,santra_hydrogen_2011,murray_prl_2012,pamuk_prl_2012}
and liquid water.\cite{laasonen1993,sprik_jcp_96,silvestrelli_jcp_99,parrinello_03,asthagiri_03,grossman_water_1,grossman_water_2,artacho2004,Kuo_04,Mcgrath_jpcA_06,Todorova_06,tuckerman_jcp_06,car_08,parrinello_09,xantheas_JCP_09,lin_jpcB_09,schmidt_jpcB_09,zhang_jctc_2011_2,ojo_cpl_2011,wang_jcp_2011,yoo_jcp_11,zhang_jctc_2011}
However, despite at least three decades of research, 
an entirely satisfactory parameter free description of energy differences between
different water clusters and ice phases has long been unattainable. 
This changed only recently, with growing compute power making it possible
to treat water clusters and solid state phases using
{\it ab initio} quantum chemical 
methods\cite{kim_jordan_1994,xantheas_1995,xantheas_JCP_02,santra_jcp_2007,santra_jcp_2009,truhlar_hexamer_2008,shields_kirschner_2008,hammond_09,jordan_10,shields_jpca_2010,bygrave_jcp_12,gillan_jcp_12,gillan_jcp_13a,delben2013jpcl,delben2013jctc}
and accurate diffusion Monte Carlo (DMC) calculations,\cite{santra_hydrogen_2011} respectively.
These calculations certainly constitute a reference for future work and are
important benchmarks for more approximate methods.

What makes the description of ice such a challenge is
that the bonding between the water molecules is determined
by fairly long range static as well as dynamic (i.e. van der Waals) dipole-dipole interactions 
and the Pauli exclusion principle between the closed shells at short distances.
Van der Waals (vdW) interactions and Pauli repulsion are difficult to handle without 
explicitly resorting to many electron techniques,
such as quantum chemical methods or
diffusion Monte Carlo. These methods, in particular
the stochastic approaches, are exceedingly demanding when 
small energy differences  between competing
phases need to be evaluated with  meV accuracy. 
The quantum chemical methods, on the other hand, are
yet restricted to small clusters, requiring fairly
complicated incremental approaches for the treatment 
of three dimensional solids or liquid phases.\cite{hermann_prl_08,manby2010chap}
Therefore currently DFT methods that are
computationally cheap, though more approximate,
are widely used. Examples for this are 
the vdW-DFT of Langreth and Lundqvist and coworkers\cite{dion2004,langreth2005,kelkkanen2009}
that employs a non-local density functional and DFT-D using
simple pair wise vdW corrections
between the constituent H and O atoms.\cite{grimme2006d2,tkatchenko2009,santra2008}
These approaches are reasonably accurate, and the verdict which
one should be preferred over the other is to some extent still debated. 

Both vdW-DFT and DFT-D approaches, however, also share the common feature that the exchange interaction,
which is an important part of the Pauli exclusion principle, is modeled
using DFT. This seems to be problematic, in particular
at high densities, where the water molecules approach each other
and the molecular charge densities start to overlap.
Indeed, both vdW-DFT and DFT-D are not entirely satisfactory in capturing
the density difference between ambient and high pressure phases.\cite{santra_hydrogen_2011}
Also it is not quite obvious whether these two approaches
are directly applicable to water on surfaces, a research area
receiving currently significant and growing attention;\cite{klimes2012,carrasco2011,carrasco2012nmat,carrasco2013} vdW-DFT is
fundamentally based on the interaction within jellium, and might work
well for ice on metallic surfaces, whereas the addition of
pair-wise atom centered vdW potentials is most likely more suitable for
water on covalent and possibly ionic substrates. 

The random phase approximation (RPA) to the correlation energy avoids these caveats, as
it combines the exact exchange with an approximate but
reasonably accurate treatment of the correlation. The exact exchange energy
is considered to be superior to semi-local exchange functionals at short
distances. The correlation part is calculated from 
the DFT based independent particle response function; for molecules, this yields fairly reliable dispersion forces at 
large distances with C$_6$ coefficients in good agreement with experiment.\cite{harl2008,lebegue2010,ren2011}
Current DFT functionals also describe the response of insulators
and metals reasonably well, and therefore, metals, semiconductors
and insulators are handled with about similar accuracy by
this approximation.\cite{harl2009,harl2010,schimka2013,yan2013}
This suggests that the RPA
should describe the interaction of water with any substrate reasonably
well. What remains to be demonstrated is whether the RPA is
accurate for the description of the intermolecular
interactions between water molecules. This is exactly
the purpose of the present work. Here we apply the Vienna Ab-initio Simulation Package (VASP)
to many phases of water ice and compare our findings with 
well established experimental data and previously calculated DMC data.
In general, we find that the RPA results compare very well with
the reference data, although the treatment of the repulsive part,
the exchange interaction, remains to some extent dependent 
on the type of orbitals used in the calculations:
with DFT orbitals, the exchange interaction is too repulsive yielding too small
binding energies and somewhat too large equilibrium lattice
constants, whereas with Hartree-Fock orbitals we find opposite trends,
too large binding energies and too small lattice constants.

\section{ Methods}

\subsection{Computational Methods}

In the present work, all calculations  were performed
using VASP.
The projector augmented wave method of  Bl\"ochl in the implementation
of Kresse and Joubert was used.\cite{blochl1994,kresse1999}
The employed potentials were constructed to conserve the scattering
properties of the atoms well up to about 20 Ry above the vacuum level. 
This was achieved by using additional projectors above the vacuum level.
Core radii of 0.95~a.u. for H and 1.5~a.u. for O were applied.
The scattering properties are, however, correctly described even
at much smaller radii of about 1.1~a.u. for O and 0.5~a.u. for H. 
Partial waves for $s$, $p$, and $d$ orbitals were included for both
O and H. Specifically,  the {\tt O$\_$GW$\_$new} and {\tt H$\_$GW} potentials as released with vasp.5.3
were used.

All plane waves with the kinetic energy lower then 800~eV were used in the DFT calculations. 
Such a larger cutoff guarantees 
convergence to a few meV in absolute energies, and similar
results could be obtained at much lower energy cutoffs. However, since
DFT calculations are computationally much less expensive than RPA calculations,
and in order to avoid tedious  convergence tests, we have chosen
this rather generous plane wave cutoff.
All DFT calculations, except when otherwise noted, are performed
using the Perdew, Burke, Ernzerhof functional.\cite{perdew1996}

Our calculations beyond DFT use two slightly different approximations.
The first one is the usual exact exchange (EXX) plus random phase approximation (RPA).
In this case, we first perform a standard PBE calculation, and then
evaluate the EXX energy using PBE orbitals and add the
correlation energy calculated in the random phase approximation with PBE
orbitals and PBE one-electron energies (EXX+RPA@PBE). 
It has been noted by Ren and coworkers\cite{ren2011} that this approximation
often underestimates the intermolecular binding energies between small molecules,
since the occupied PBE orbitals are spatially too delocalized. This 
results in a too strong Pauli repulsion at the equilibrium distance
and, resultantly, too large intermolecular bond lengths. To resolve
this issue, various approximations have been suggested among
them replacing the exact exchange evaluated employing PBE orbitals
by the Hartree-Fock exchange,\cite{ren2011} or a restricted summation of the singles
contributions in diagrammatic perturbation theory (rSE).\cite{ren2011,paier2012,ren2012,ren_renormalized_2013}
We believe that none of these solutions is entirely satisfactory, as all of them
assume in essence that the one-particle reduced density matrix $\gamma({\bf r}, {\bf r}')$ 
from Hartree-Fock   is more accurate than the reduced density matrix from 
the approximate density functional. This might be true for some specific cases, 
such as small molecules,  large band gap insulators, or regions far from any atomic core, however, 
it can hardly be true
for metals or small gap insulators, where present density functionals are far
more accurate than the Hartree-Fock approximation. 
It is, however, clear that the Hartree-Fock orbitals are spatially more contracted
than the PBE orbitals, and evaluation of the exact exchange energy with those
orbitals hence reduces the intermolecular  Pauli repulsion. Consequently the intermolecular
distances become smaller, and past experience suggests that this often improves agreement 
with experiment.\cite{ren2011}
To obtain a--- what we believe ---lower bound for the lattice constants, we have therefore
also evaluated the exact exchange energy using Hartree-Fock orbitals. 
Even in this case, the correlation energy is calculated with PBE orbitals
and PBE one electron energies. In this work, we refer to this scheme as
HF+RPA@PBE.

The EXX+RPA@PBE and the HF+RPA@PBE  calculations are
performed at a more modest computational setup than the PBE calculations
discussed above. The plane wave
energy cutoff for the orbitals was set to $E^{\rm PW}_{\rm cut } = 450$~eV. When summations
over unoccupied Kohn-Sham states are required (virtual orbitals), 
all orbitals spanned by the basis set are determined by exact
diagonalization of the Kohn-Sham Hamiltonian. The correlation
energy in the random phase approximation is then calculated in 
the usual manner as
\begin{eqnarray}\label{eq:ece}
E^{\rm RPA} = \int_{0}^{\infty} \frac{d\omega}{2\pi} \, {\mathrm{Tr}}\{\ln\left[1-\chi^{\rm KS}(i\omega) \nu
 \right] + \chi^{\rm KS}(i\omega) \nu\}\,,
\end{eqnarray}
where $\chi^{\rm KS}$ is the independent particle response function evaluated
using PBE orbitals and one electron energies, and $\nu$ is the
Coulomb kernel. The response function
itself is also expanded in a plane wave basis set. The plane wave
cutoff for this basis set is set to 210--300 eV (smaller than the basis set for
the orbitals), and the correlation energy is extrapolated to the infinite basis set limit, assuming
that the basis set error falls off like the inverse of the number of plane waves included
in the basis set for the response function.\cite{harl2008} 
In the VASP code, this requires a single calculation and the extrapolation 
is performed automatically by the code, requiring a minimum of extra compute time.
The structures used for the RPA calculations were  determined
by completely relaxing all internal parameters of the structures (including the cell shape) 
at a set of volumes employing the PBE functional (and the previously mentioned cutoff of 800~eV).
In the subsequent RPA calculations, the PBE structures were kept fixed since
forces and the stress-tensor are presently not available within the RPA.
Similar strategies are also routinely adopted in DMC simulations and
most coupled cluster quantum chemistry calculations.

For ice, the RPA energy volume curves usually span a very small energy
range of the order of 10~meV per molecule, when the volume varies by 10~\%.
These small energy changes make converged calculations particularly challenging. 
For instance, when the volume changes, the number of plane waves {\bf  G}
 at a $\bf q$ point in the Brillouin zone changes disruptively with volume:
\[
 \frac{\hbar^2 |{\bf  G}+{\bf q}|^2}{2 m_e} <  E_{\rm cut }.
\]
This problem is more severe for high symmetry structures than for low
symmetry structures, since reciprocal lattice shells show more degeneracies
in high symmetry structures.
The smoothness can be improved by either increasing the cutoffs (which we found
unpractical for  the present calculations) or increasing the number of sampling points $\bf q$
in the Brillouin zone. In this work, we have
followed the second approach, {\it i.e.} increasing the number of $\bf q$ points
until a smooth energy-volume curve was obtained. For some
phases, a sizable residual noise, however, remains, and
the bulk moduli might exhibit error bars of about 10--20~\%
(estimated from different $\bf q$ point sets).

To obtain the ice binding energies with respect to the free water
molecule, accurate reference values for the exchange and correlation
energy of the water molecule are required. The exchange energies, EXX and HF, 
were determined by calculating the energy of a single H$_2$O molecule
in a box with the box size systematically varied between 7 and 21~\AA.
The correlation energy was determined in a smaller  7~\AA\ box using
$3\times 3 \times  3$ k-points.
Although absolute energies are not directly transferable between different codes or potentials
we also report the final molecular results obtained by VASP. The final RPA correlation energy of a single water molecule  is 
$-12.426$~eV, the EXX energy using PBE orbitals and the HF energy are $-29.254$~eV and $-29.479$~eV, respectively.
As a matter of fact, the (self-consistently evaluated) HF energy is significantly lower than the
EXX energy.

\subsection{Considered Phases}

\begin{table*}[t]
\caption{
Summary of ice phases considered in the present work. The experimental Bravais lattice (exp),
as well as the considered approximation to the experimental structure (calc) are
specified. The total number of molecules is given in the column ``mol''.
The column ``k-points'' indicates the number of divisions ($n$) in each reciprocal lattice direction
used to generate a uniform k-point grid (e.g. $n \times n \times n$).
 }
\label{tab:icestructures}
\begin{ruledtabular}
\begin{tabular}{lccccccc}
phase &  space group & exp & calc & mol. & k-points\\
\hline
 I$_h$  &    P6$_3$/mmc (\#194) &hexagonal    & simple monoclinic &          12 & 3 \\
 XI Cmc2$_1$  & Cmc2$_1$ (\#36) &orthorhombic &  base c. orthorhombic &      4  & 6 \\
 XI Pna2$_1$  & Pna2$_1$ (\#33) &orthorhombic &simple orthorhombic &     8  & 3 \\
 I$_c$(a)  & I4$_1$md (\#109)   &tetragonal  & body c. tetragonal &       2  & 7 \\
 I$_c$(b)  & Pna2$_1$ (\#33)   &orthorhombic & simple orthorhombic &      4  & 5 \\
 I$_c$(c)  & P4$_1$ (\#76)   &tetragonal  & simple tetragonal &           8  & 4 \\
 I$_c$(d)  & P4$_1$2$_1$2 (\#92)   &tetragonal & simple tetragonal &      4  & 4 \\
 IX  &      P4$_1$2$_1$2 (\#92) & tetragonal &simple tetragonal &        12  & 3 \\
 II  &      R$\bar{3}$ (\#148) & rhombohedral &trigonal (rhombohedral) & 12  & 3 \\
 XIII  &    P2$_1$/a (\#14)  & monoclinic&simple monoclinic &  28 &  1  \\
 XV&          P1 (\#1)  &triclinic &triclinic &       10  &   3 \\
 VIII  &       I4$_1$/amd (\#141) & tetragonal &body c. tetragonal &      4  &  6 \\
\end{tabular}
\end{ruledtabular}
\end{table*}

\begin{figure}[t]
 \centering
 \includegraphics[width=70mm]{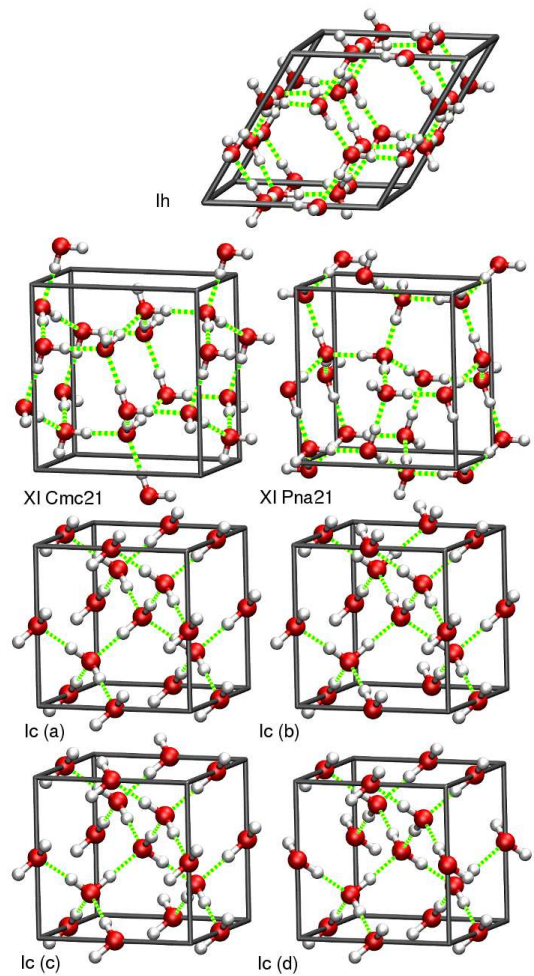}
 \caption{Low density structures of ice considered in the present work.
 Large red and small white spheres indicate the oxygen and hydrogen atoms, respectively.
Dashed (green) lines indicate hydrogen bonds. Full thick lines demarcate the unit cell. 
 \label{fig:ice1}}
\end{figure}

\begin{figure}[t]
 \centering
 \includegraphics[width=75mm]{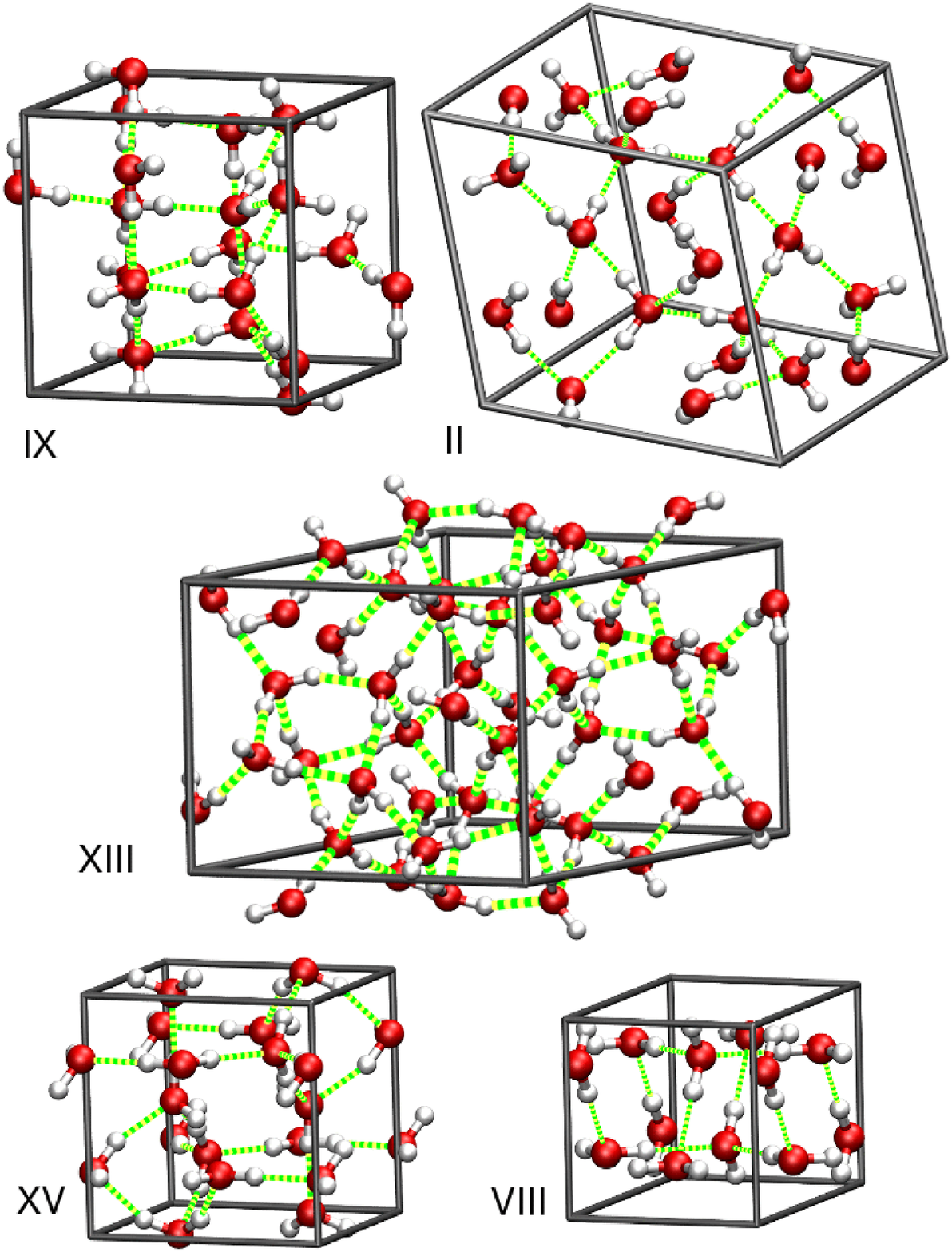}
 \caption{High density structures of ice considered in the present work.
 Color coding is the same as in figure~\ref{fig:ice1}.
 \label{fig:ice2}}
\end{figure}

From the known crystalline phases of ice we have considered the ones listed in  Table~\ref{tab:icestructures} and 
depicted in figures~\ref{fig:ice1} and~\ref{fig:ice2}. All structures follow the usual building rules for ice: 
each water molecule has four nearest neighbor water molecules, donates two hydrogen bonds, and accepts two hydrogen bonds. 
The hydrogen atoms (protons) are typically not fixed in their high-symmetry positions. This results in the formation of 
so-called proton disordered phases, {\it i.e.} possible configurations of proton positions consistent with the ice rules.
The phases of ice under scrutiny in the present work can be arranged into two groups depending on their densities. (i) Low-density phases 
(see Fig.~\ref{fig:ice1}): proton disordered hexagonal ice (I$_h$), the most stable proton ordered form of I$_h$ (XI Cmc2$_1$),
the second most stable proton ordered phase of I$_h$ (XI Pna2$_1$), and four proton ordered cubic phases (I$_c$ a-d).
(ii) High-density phases: the proton ordered form of ice III (ice IX), the proton ordered ice II, the proton ordered form 
of ice V (ice XIII), the proton ordered form of ice VI (ice XV), and the proton ordered form of ice VII (ice VIII).
In the following, we will briefly summarize the structural details of each phase.

Hexagonal ice (I$_h$) is the common phase of ice naturally encountered. 
The protons are usually disordered, although partial order can be induced
by careful, slow annealing in the presence of some ionic ``catalyst" such as 
KOH,\cite{leadbetter_equilibrium_1985, howe_determination_1989}
which increases the usually slow reorientation rate in ice. 
To model the disorder, we have adopted the scheme
used by Hamann.\cite{hamann_h_2o_1997} This 
lowers the symmetry to  monoclinic with a structural
model containing 12 molecules. All our energies are referenced to 
this structure in the final figures.
This needs to be considered when comparison is 
made with other calculations. However, a recent study of Santra~{\it et al.} suggests
that the energy difference to a larger 96 molecule cell is only 
in the range of 1~meV.\cite{santra_accuracy_2013} 
To study proton ordered variants of hexagonal ice, we have included 
one hexagonal structure with ferroelectric and one with anti-ferroelectric
order, respectively. The ferroelectrically ordered phase is realized in ice XI
with space group Cmc2$_1$. The structure was first resolved experimentally
by Leadbetter~{\it et al.}\cite{leadbetter_equilibrium_1985} showing ``polar" order
on a length scale of about 40~\AA.
An anti-ferroelectrically ordered phase can be realized in the space group  Pna2$_1$ and was
suggested by Davidson et al.\cite{davidson_proposed_1984} Further prototypically
ordered structures have been studied systematically in Refs.~\onlinecite{hirsch_quantum-chemical_2004, Profeta}
using density functional theory and a plane wave code.

Cubic ice (I$_c$) is experimentally difficult to prepare and forms only
under certain conditions,  for instance, it is believed to form in the Earth's upper troposphere at
temperatures of less then 220 K.\cite{murray_formation_2005}
Above 240 K, cubic ice tends to transform to hexagonal ice. 
As for hexagonal ice, cubic ice is usually proton disordered.
Here we considered four different proton ordered phases out of the
11 different proton configurations enumerated in Ref.~\onlinecite{raza_proton_2011}.
The first phase is fully ferroelectrically ordered and denoted as I$_c$(a),
whereas the other three are ferrielectrically (b-c) and anti-ferroelectrically (d) ordered; the number
of anti-ferroelectrically ordered neighbors increases as one goes from structure
(b) to structure (d). 
Details of the structures are discussed elsewhere.\cite{geiger2013}

Ice IX is stable at temperatures below 140 K and pressures between
300 and 400~MPa. Experimentally it is
formed by slow cooling of tetragonal crystalline ice III (formed by cooling water down 
to 250 K at 300 MPa).\cite{whalley_ice_1968} 
Both disordered ice III and ordered ice IX  have the space group P4$_1$2$_1$2 (\#92).
At similar pressures ($300$~MPa) and low temperature ($198$~K), ice II (the rhombohedral crystalline form of ice)
can be formed from ice I$_h$.\cite{kamb_ice._1964}
Ice II is a proton ordered phase with space group R$\bar{3}$ (\#148),
with no direct proton disordered counterpart.\cite{chaplin_water_2012}
The structure is  characterized by two hexagonal rings, connected by hydrogen bonds. 

The monoclinic proton-ordered ice XIII was successfully prepared and 
structurally determined by neutron powder diffraction in 2006.\cite{salzmann_preparation_2006}
After doping with HCl, it can be formed from the corresponding proton-disordered monoclinic phase of ice (ice V)  
at temperatures slightly below $130$~K and applying a pressure of $0.5$~GPa.
In the pressure range from $0.8$~GPa to $1.5$~GPa and at similar low temperatures,
the proton-ordered counterpart to the proton-disordered phase VI has been identified in 2009\cite{salzmann_ice_2009} 
and named ice XV (triclinic).  

The final highest pressure phase considered here is
ice VIII. It is the proton ordered form of the proton disordered ice VII with tetragonal space 
group I4$_1$/amd (\#141).  The oxygen sublayer is the same for both structures and all molecules 
have an equivalent environment. 
The structure consists of two inter-penetrating, but not interconnected cubic ice I$_c$ sublattices.\cite{kuhs_structure_1984}
The sublattices have opposite dipole moments resulting in an anti-ferroelectric ordering.\cite{pruzan_stability_1993}

\section{Results}

In the first subsection we present results for few selected
phases, elaborating on illustrative tests and 
important issues. In the subsequent subsections, binding energies 
and equilibrium volumes are discussed.

\subsection{Preliminary Remarks}

\begin{figure}
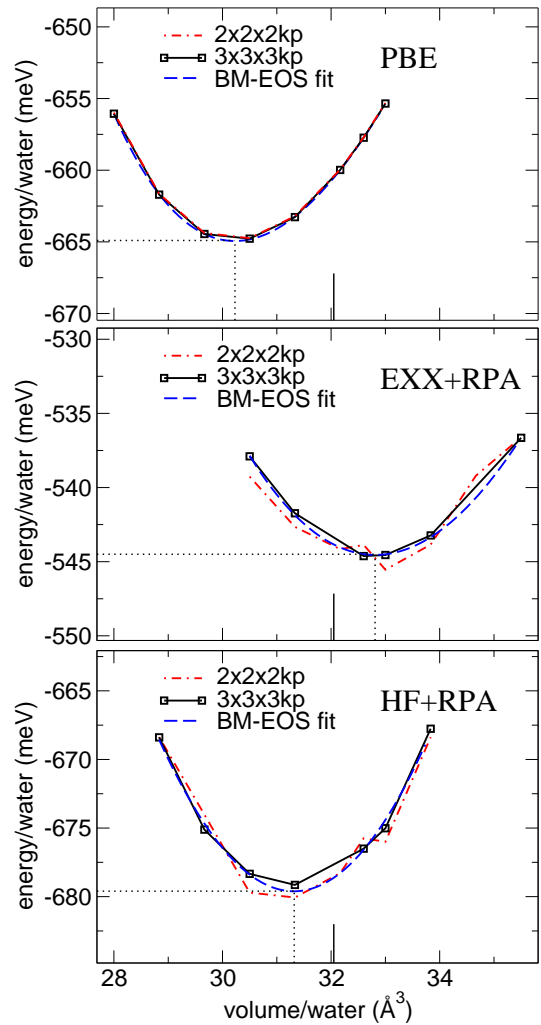

\centerline{\includegraphics[width=7cm,clip=true]{figure3a}}
\centerline{\includegraphics[width=7cm,clip=true]{figure3b}}
\centerline{\includegraphics[width=7cm,clip=true]{figure3c}}
\caption{Binding energy per water molecule versus volume  for 
ice I$_h$ (P6$_3$/mmc) for PBE, EXX+RPA@PBE, and
 HF+RPA@PBE. The experimental volume is shown as a vertical line. 
The dotted lines indicate the theoretical volume and energy.
\label{fig:iceh}
}
\end{figure}

\subsubsection{Optimized volume} 
\begin{figure}
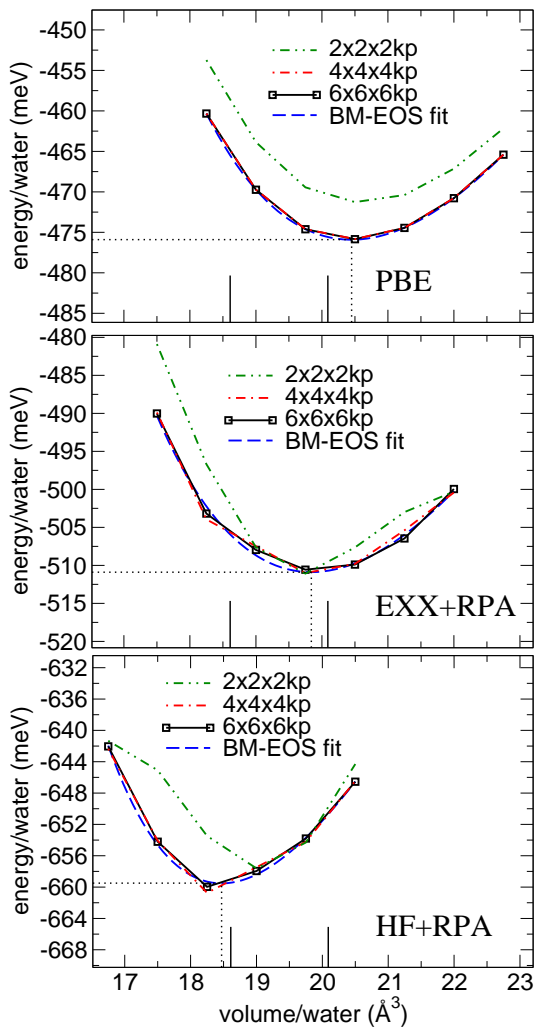

\centerline{\includegraphics[width=7cm,clip=true]{figure4a}}
\centerline{\includegraphics[width=7cm,clip=true]{figure4b}}
\centerline{\includegraphics[width=7cm,clip=true]{figure4c}}
\caption{Binding energy per water molecule versus volume  for 
ice VIII  for PBE, EXX+RPA@PBE, and
HF+RPA@PBE. The experimental volume is shown as a vertical line.
\label{fig:iceviii}
}
\end{figure}

To illustrate the general behavior and, specifically, 
the rather slow k-point convergence, we  
show in Figs.~\ref{fig:iceh} and~\ref{fig:iceviii} the energy as a function 
of the cell volume per water molecule for different k-point meshes.
The data computed within PBE, EXX+RPA@PBE and HF+RPA@PBE
are shown for ice I$_h$ (low density case) in Fig.~\ref{fig:iceh} and for ice VIII (high density case)
in Fig.~\ref{fig:iceviii}.

The first thing to notice is that, for ice I$_h$ at the PBE level, the equilibrium volume is
much too small compared to experiment. 
Explicitly including many-body correlation effects in diagrammatic perturbation theory
clearly improves upon this point, with the equilibrium volumes now
approaching the experimental values.
For the high density ice VIII phase, 
the different available experimental values for the volume  
(18.61~\AA$^3$, and 20.09~\AA$^3$) \cite{jorgensen_structure_1984, yoshimura_high-pressure_2006}
hamper a definite assessment of the various theoretical methods. 
In this case, PBE certainly yields much too large volumes. 
This can be attributed to the lack of vdW interactions that become more important
at small volumes, where the number of neighbors increases.\cite{santra_hydrogen_2011}
Also in this case, the RPA delivers an improved description compared to PBE.
The EXX+RPA optimized volume is bracketed by the experimental values, whereas the
HF+RPA estimate is much closer to the lower experimental value.
Following the expectations already outlined in the
introduction, the HF+RPA@PBE equilibrium volumes are smaller 
than the EXX+RPA@PBE volumes. 
In fact, both, HF+RPA@PBE and EXX+RPA@PBE roughly bracket
the experimental volume, something we will also observe for
other ice phases.

\subsubsection{Convergence with k-points} 

As for the k-point convergence, we find that the DFT results converge rapidly with k-points, and show only 
very little jaggedness, whereas the RPA results exhibit quite some residual
noise for a coarse $2\times2\times2$ k-point grid, especially for the high density
phase (ice VIII). The jaggedness is mostly gone in the low density I$_h$ phase
with $3\times3\times3$ points, whereas in ice VIII reasonable smoothness is only achieved 
from $4 \times 4 \times 4$ k-points on. For the converged k-point grids,
the data can be well fitted with a 3rd order Birch-Murnaghan equation of state.

The final k-point grids for each phase are summarized in 
Table \ref{tab:icestructures}. As already mentioned, in some
cases a residual  jaggedness prevails, since we were unable to increase
the k-point grids significantly beyond the values in the Tables with the present code.
In any considered case, however, the
minimum was clearly resolved in the energy volume curve.

\subsubsection{ Structure of ice IX} 

Ice IX has a tetragonal structure, and initially we performed
a structural optimization for all internal parameters (including
the $c/a$ ratio) at each volume using the PBE functional.
This lead to results in disagreement with the previous calculations.\cite{santra_hydrogen_2011}
However, the previously reported calculations were performed
with the cell shape fixed to the experimentally determined
structure. Relaxation of the lattice shape changes
the results only little for all, but the ice IX structure.
In fact, PBE yields only a mediocre description of the $c/a$
ratio for ice IX, with the value approaching 1.123
whereas the experimental values are close to 1.01.\cite{la_placa_nearly_1973,londono_neutron_1993}
In this case, vdW interactions are particularly important
along the $c$ direction, and only inclusion of them
improves the $c/a$ ratio. For instance DFT-D yields a $c/a$ ratio around 1.03.
To maintain compatibility
with the rest of the results, we decided to stick to
the optimization of the structures using the PBE functional,
but kept the $c/a$ ratio of ice IX fixed to the experimental value. Both results, 
the one with the full optimization of the $c/a$ ratio and the one with the $c/a$ ratio
fixed to the experimental values are shown in the tables,
whereas the figures report on the results for fixed $c/a$ ratio.
As Tab. \ref{tab:iceenergies} shows, the RPA yields a lower energy with
the $c/a$ ratio fixed to the experimental value.

\subsection{Binding energies}

\begin{table}[!htb]
\caption{Binding energies in eV per water molecule as obtained for different phases of 
ice with PBE, EXX+RPA@PBE, and HF+RPA@PBE.
The energies were evaluated at the equilibrium volumes corresponding to 
a given method and phase. 
}
\label{tab:iceenergies}
\begin{ruledtabular}
\begin{tabular}{llll}
Phase       & PBE          & EXX + RPA$^{\rm EXX}$ & HF + RPA$^{\rm HF}$ \\
\hline
I$_h$       &   $-$0.6649 &       $-$0.5445 &       $-$0.6796\\
XI Cmc2$_1$ &   $-$0.6678 &       $-$0.5476 &       $-$0.6831\\
XI Pna2$_1$ &   $-$0.6637 &       $-$0.5443 &       $-$0.6779\\
I$_c$(a)    &   $-$0.6682 &       $-$0.5478 &       $-$0.6824\\
I$_c$(b)    &   $-$0.6654 &       $-$0.5448 &       $-$0.6793\\
I$_c$(c)    &   $-$0.6641 &       $-$0.5417 &       $-$0.6769\\
I$_c$(d)    &   $-$0.6627 &       $-$0.5425 &       $-$0.6773\\
IX          &   $-$0.6199 &       $-$0.5290  &       $-$0.6652\\
IX c/a exp  &   $-$0.6126 &       $-$0.5362 &       $-$0.6735\\
II          &   $-$0.5918 &       $-$0.5358 &       $-$0.6712\\
XV          &   $-$0.5474 &       $-$0.5223 &       $-$0.6620 \\
VIII        &   $-$0.4759 &       $-$0.5111 &       $-$0.6593\\
\end{tabular}
\end{ruledtabular}
\end{table}

\begin{figure}
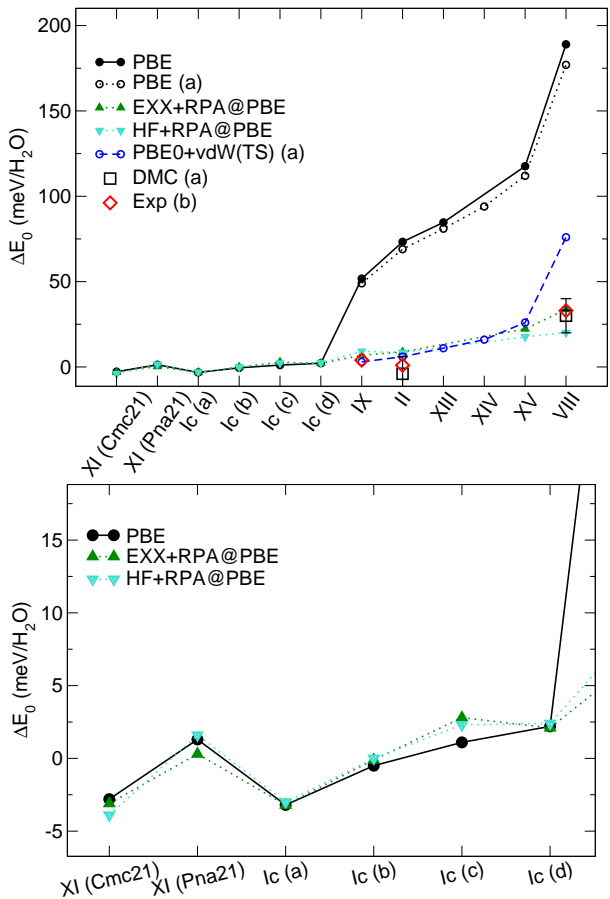

\centerline{
\includegraphics[width=8cm,clip=true]{figure5a}}
\centerline{\includegraphics[width=7.8cm,clip=true]{figure5b}}
\caption{Binding energies per water molecule relative to ice I$_h$. $\Delta E_0$ is the energy with respect to  
proton disordered I$_h$ for the same method: $\Delta E_0 = E_0({\rm phase}) - E_0({\rm I}_h)$. 
Other theoretical data marked with (a) are from Ref.~\onlinecite{santra_accuracy_2013} [PBE, PBE0-vdW(TS), DMC].
Experimental data (b) are from  Ref.~\onlinecite{whalley_energies_1984}.
\label{fig:iceenergies}
}
\end{figure}

The binding energies of all considered phases are summarized
in Tab.~\ref{tab:iceenergies} and Fig.~\ref{fig:iceenergies}.
In the figure, the energies are shown with respect to ice $I_h$.
Let us first concentrate on the phases at ambient pressure,
hexagonal ice and cubic ice (lower panel in Fig.~\ref{fig:iceenergies}). 
From the figure, it is clear that
there is very little difference between the standard density functional
theory calculations (PBE) and the two RPA variants for relative energies. 
Note that, mostly because of the insufficient sampling of the Brillouin zone, 
the errors in the RPA energies are about 1~meV for relative energies. 
Clearly the proton disordered hexagonal
phase is about 3 meV higher in energy than the ferroelectrically
ordered phase XI Cmc2$_1$. 
We note that this is in agreement
with the experimental data that also predict a long range
ferroelectrically ordered phase as the actual low temperature ground state 
structure of hexagonal ice.\cite{leadbetter_equilibrium_1985}
The anti-ferroelectric phase considered here (XI Pna2$_1$) is slightly
higher in energy than the disordered phase of Hamann, but the energy difference
is as small as 2 meV in PBE and  HF+RPA@PBE, and vanishing within the error bars for
EXX+RPA@PBE. One would expect that the cubic phase also
prefers a ferroelectric order.  Indeed, this is confirmed, 
with the  I$_c$(a)  being practically isoenergetic with the hexagonal
ferroelectric phase. Along the sequence I$_c$(a) $\rightarrow$
I$_c$(d)  the number of anti-ferroelectrically ordered
neighbors increases. The energy increase is about equal for PBE
and the two RPA variants, with a somewhat steeper increase in the RPA until I$_c$(c),
and almost isoenergetic results for I$_c$(c) and I$_c$(d) in the RPA.
The important observation is that for the results at ambient
pressure, PBE seems to capture all essential trends, and
the RPA yields qualitatively and even quantitatively the same results. This is most likely related to the
fact that the energy ordering is determined by the long
range dipole-dipole interactions between different 
water molecules.

It is worthwhile mentioning that simple electrostatic non-polarizable water models predict 
that the anti-ferroelectric phase is more stable than
the ferroelectric phase,\cite{Lekner1997} whereas DFT and more accurate methods predict that the
ferroelectric order is preferred, both for the hexagonal as
well as for the cubic phase. This can be related
to the polarizability and additionally induced dipoles on the water molecules,
which are neglected in simple rigid electrostatic water models.

The precision of PBE deteriorates quite dramatically, as one moves to 
high pressure phases. 
Pressure induces a 
sizable increase in the density and a reduction of the equilibrium volume.
The energy ordering is shown in the top panel of
Fig.~\ref{fig:iceenergies}. 
Let us start with a comparison between the present PBE calculations and 
previously published data. 
Generally our data reproduce the previous trends quite well, 
even though our data points tend to lie at slightly higher energies than
the previous calculations.
For ice VIII, the deviation amounts to about 20~meV.
The reason for this discrepancy is that the present calculations
were performed with a fairly large core oxygen PAW potential
to reduce the computational cost for the subsequent RPA calculations. 
In any case, it is clear that compared to experiment 
both PBE calculations show a much too steep increase of the energy as the volumes decrease.

The RPA energy differences with respect to ice I$_h$ agree very well with the previously
published DMC data for ice II and ice VIII. 
Specifically, ice II becomes stabilized
by almost 60 meV compared to  PBE, 
and the stabilization is even more dramatic for ice VIII
where the energy difference decreases by almost 170 meV. 
There is a slight difference between the EXX+RPA@PBE
and HF+RPA@PBE as the pressure increases: as already argued before, HF 
leads to a contraction of the density and therefore
reduces the Pauli repulsion. Hence, the increase in
the energy is smaller with HF+RPA@PBE than with EXX+RPA@PBE.

In our present calculations ice IX and ice II are
almost iso-energetic lying about 10~meV above the
hexagonal disordered ice phase.
We believe that this result is very reasonable and
consistent with the experimental situation that
both phases can be prepared at similar preparation
conditions and pressures from different parent ice phases.
The experimental estimates see ice II at slightly lower energies,
but we can not confirm this (neither do the simulations
using vdW corrected functionals).
In the DMC the ice II is at lower energies than
the disordered hexagonal phase, which is not confirmed by our RPA data, either.
However, considering the statistical error bars of the DMC calculations, we believe that this 
residual difference is not really meaningful.

For ice XV, our predicted energy differences agree very well with
vdW corrected DFT calculations. However, for  ice VIII, the increase in the energy is
too steep using the vdW(TS) corrections. 
Here, the RPA does clearly better and yields energy changes in
very good agreement with DMC simulations and experiment.

With respect to isolated water molecules, the RPA energies are not
on par with the DMC results. 
From Table~\ref{tab:iceenergies},
we find for ice I$_h$ a binding energy of $-545$~meV 
and $-680$~meV for EXX+RPA@PBE  and  HF+RPA@PBE, respectively.
The experimental and DMC values are $-610$ and $-605\pm$5~meV, respectively.
Clearly the EXX+RPA underbinds by about 65 meV, whereas the
HF+RPA overbinds by about the same amount. 
The correct value lies almost exactly in between the
two values, suggesting that, in this case, a mean
field description with 50~\% Hartree-Fock exchange and 50~\% EXX
evaluated using DFT orbitals would yield very good results. 
This is a sensible result: in H$_2$O and ice, the screening
is very weak, which implies that hybrid functionals should
include more exchange than in the typical hybrid functionals such
as PBE0\cite{adamo1999} and HSE.\cite{heyd2003hse,heyd2006hse} 
The qualitatively ``best" orbitals
are most likely obtained by a half-half hybrid functional, with ``best"
implying that such a functional yields a one-particle density matrix
very close to the true one-particle density matrix.

\subsection{Equilibrium volumes}

\begin{table*}
\begin{threeparttable}
\caption{Equilibrium volumes in \AA$^3$/molecule of the considered phases for
PBE, EXX+RPA@PBE, HF+RPA@PBE, and experiment.
}
\label{tab:icevolume}
\begin{ruledtabular}
\begin{tabular}{lrrrrrr}
Phase       & PBE  &  PBE0+vdW & EXX+RPA@PBE & HF+RPA@PBE  & DMC &    EXP  \\
\hline
I$_h$       & 30.23  & 29.88 & 32.81 & 31.31 & 31.69 & 32.05$^a$, 32.50$^b$ \\
XI Cmc2$_1$ & 30.33  &       & 32.43 & 31.03 &       & 32.15$^c$, 31.92$^d$, 31.99$^e$ \\
XI Pna2$_1$ & 30.23  &       & 32.78 & 31.38 &                      \\ 
I$_c$(a)    & 30.20  &       & 32.77 & 31.53 &       &  32.105$^b$ \\ 
I$_c$(b)    & 30.20  &       & 32.67 & 31.20 &  \\ 
I$_c$(c)    & 30.21  &       & 32.64 & 30.87 &  \\ 
I$_c$(d)    & 30.23  &       & 32.57 & 31.30 &  \\ 
IX          & 26.75  &       & 27.32 & 25.60 &       & 25.80$^f$, 25.63$^g$\\ 
IX c/a exp  & 25.66  & 23.85 & 26.61 & 24.98 &       & 25.80$^f$, 25.63$^g$\\ 
II          & 24.63  & 23.63 & 25.14 & 23.76 & 24.7  & 24.97$^h$, 24.63$^i$ \\ 
XIII        & 23.67  & 22.47 &       &       &       & 23.91$^j$ \\ 
XV          & 22.45  & 21.45 & 22.48 & 21.32 &       & 22.53$^k$ \\ 
VIII        & 20.45  & 19.70 & 19.84 & 18.47 & 19.46 & 18.61$^l$, 20.09$^m$ \\ 
\end{tabular}
\end{ruledtabular}
\begin{tablenotes}
\item[a] Ref. \onlinecite{hobbs_ice_1974}\\
\item[b] Ref. \onlinecite{vega_radial_2005}\\
\item[c] Ref. \onlinecite{leadbetter_equilibrium_1985}\\
\item[d] Ref. \onlinecite{howe_determination_1989}\\
\item[e] Ref. \onlinecite{line_high_1996}\\
\item[f] Ref. \onlinecite{la_placa_nearly_1973}\\
\item[g] Ref. \onlinecite{londono_neutron_1993}\\
\item[h] Ref. \onlinecite{fortes_incompressibility_2005}\\
\item[i] Ref. \onlinecite{lobban_pt_2002}\\
\item[j] Ref. \onlinecite{salzmann_preparation_2006}\\
\item[k] Ref. \onlinecite{salzmann_ice_2009}\\
\item[l] Ref. \onlinecite{jorgensen_structure_1984}\\
\item[m] Ref. \onlinecite{yoshimura_high-pressure_2006}\\
\end{tablenotes}
\end{threeparttable}
\end{table*}

\begin{figure}
\includegraphics[width=8cm,clip=true]{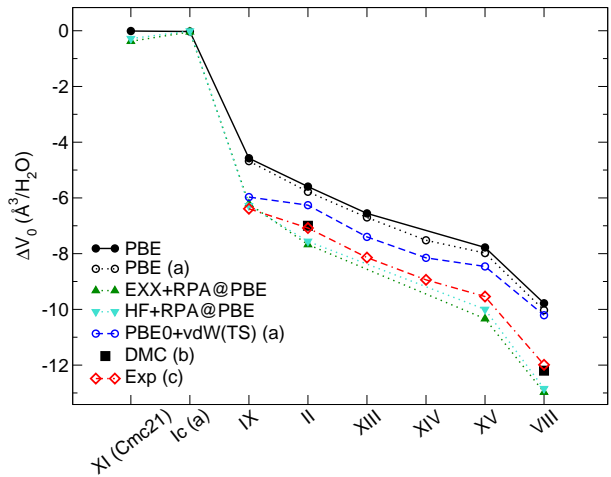}
\caption{Equilibrium volumes relative to ice I$_h$. $\Delta V_0$ is the volume with respect to the I$_h$ volume for the same method: $\Delta V_0 = V_0({\rm phase}) - V_0({\rm I}_h)$.
Other theoretical data marked with (a) and (b) are from Ref.~\onlinecite{santra_accuracy_2013} [PBE, PBE0-vdW(TS), DMC].
Experimental data (c) are compiled in Table~\ref{tab:icevolume}.}
\label{fig:icevolume}
\end{figure}

The equilibrium volumes are summarized in Tab.~\ref{tab:icevolume} and Fig.~\ref{fig:icevolume}.
As expected from the energies, Tab.~\ref{tab:icevolume}  suggests that 
EXX+RPA@PBE again underbinds (too large volumes), whereas
HF+RPA@PBE overbinds (too small volumes).  As before, a mean field description with 
50~\% Hartree-Fock exchange and 50~\% EXX employing DFT orbitals would yield very good volumes,
confirming the final conjecture of the previous section.
To delineate this constant volume error, we show in Fig.~\ref{fig:icevolume}
the relative change of the volume with respect to disordered hexagonal
ice I$_h$. The problem of PBE is that the volume changes too little
for the high pressure phase; from hexagonal or cubic ice to ice VIII only by 10~\AA$^3$
per water molecule, whereas the experimental value is closer to 12~\AA$^3$.
Van der Waals corrections improve upon this value, but only slightly.
In fact, they reduce all volumes by roughly the same fraction, 
so that changes in the relative volumes $\Delta V_0$ differ insignificantly
between PBE and PBE-D.

The only method that yields almost exact results is DMC, but
at a very steep compute cost.
RPA exhibits a very good performance with little differences in
the volume variations between EXX+RPA@PBE and HF+RPA@PBE.
On first sight, the ``compressibility" seems to be somewhat overestimated, 
with the volume changes being about 1~\AA$^3$ too large for ice VIII. 
However, our data have not been corrected for zero point energy effects.
Accounting for them increases the volumes and more so
for the high pressure phases. 
The zero point expansion is about 5~\% stronger
for ice VIII compared to ice I$_h$,\cite{murray_prl_2012,santra_accuracy_2013}
corresponding to roughly 1~\AA$^3$. 
Including the zero point vibration effects will hence give almost perfect agreement 
in the volume changes for ice VIII with experiment.

\section{Summary and Conclusions}

The present work indicates that the random phase approximation
yields a satisfactory description of ice for both the densities 
as well as the relative energies  (compare Fig. \ref{fig:iceenergies}). 
In principle, such a good description is not astonishing, since the random
phase approximation seems to capture both van der Waals interactions
as well as covalent bonding contributions reasonably well.
Actually, when only the second order contribution is taken
into account, the RPA reduces to the direct term in second order M{\o}eller Plesset (MP2)
perturbation theory, and MP2 is known to work very well for
energy differences in water clusters.

From our point of view, RPA also improves upon simple pair-wise
interaction potentials or vdW density functional theory. 
These two methods decrease the equilibrium volumes of ice compared to
standard semi-local functionals by
roughly the same amount for low and high density phases.\cite{santra_accuracy_2013}
As a result, the volume change from hexagonal ice
to ice VIII is too small. 
The RPA almost entirely mends this error (compare Fig. \ref{fig:icevolume}), in particular,
when  zero point vibration corrections are accounted for.

Up to date, only  diffusion Monte Carlo  was
able to attain a similar--- or more precisely ---slightly superior description, but
as already emphasized, at a very steep computational cost.
RPA achieves results that are close to DMC, but at a fraction of the computational cost.
For instance, the total compute time for the rather complicated
ice XIII phase (28 molecules) is about 4 hours on 64 cores,
and the compute time for ice IX (12 molecules) about 2 hours on 16 cores
(both calculations performed using $2\times 2  \times 2$ k-points).
These favorable timings were achieved with a not yet released 
RPA code that scales cubically with system size and
linearly with the number of k-points.
Given the favorable results for the energetics obtained
here and these favourable timings, we believe an {\it ab initio} treatment of ice 
on surfaces is now within reach.
The only downside of the random phase approximation  is that the
ice binding energies with respect to isolated water molecules
are in error by about 50~meV. 

In the present work, we have used two RPA flavors: 
the usual combination of EXX+RPA@PBE, where both the RPA correlation
and the  EXX are evaluated using the same DFT orbitals, and
the combination of RPA@PBE with exact Hartree-Fock energies.
The first method underestimates the binding energies and overestimates
all predicted volumes. The second approximation overestimates the binding energy
and underestimates the equilibrium volumes. A simple
solution to this problem is to determine the exact exchange energy from 
orbitals obtained with a so called half-half functional, a functional
where half the HF exchange and half the density functional theory
exchange is used. 
The final results are merely in-between the two
limiting approximations considered here and in almost perfect agreement
with the experimental and DMC results.

{\em Acknowledgment:}
This work was supported by the Austrian Science Fund (FWF)  within the SFB ViCoM (Grant F 41).
We thank Biswajit Santra for providing us with the ice structures. 
C.F. and M.M. thank Gianni Profeta for providing useful structural information on ice XI.
Supercomputing time on the Vienna Scientific cluster (VSC) is gratefully acknowledged.
\bibliography{master}

\begin{thebibliography}{100}%
\makeatletter
\providecommand \@ifxundefined [1]{%
 \ifx #1\undefined \expandafter \@firstoftwo
 \else \expandafter \@secondoftwo
\fi
}%
\providecommand \@ifnum [1]{%
 \ifnum #1\expandafter \@firstoftwo
 \else \expandafter \@secondoftwo
\fi
}%
\providecommand \enquote [1]{``#1''}%
\providecommand \bibnamefont  [1]{#1}%
\providecommand \bibfnamefont [1]{#1}%
\providecommand \citenamefont [1]{#1}%
\providecommand\href[0]{\@sanitize\@href}%
\providecommand\@href[1]{\endgroup\@@startlink{#1}\endgroup\@@href}%
\providecommand\@@href[1]{#1\@@endlink}%
\providecommand \@sanitize [0]{\begingroup\catcode`\&12\catcode`\#12\relax}%
\@ifxundefined \pdfoutput {\@firstoftwo}{%
 \@ifnum{\z@=\pdfoutput}{\@firstoftwo}{\@secondoftwo}%
}{%
 \providecommand\@@startlink[1]{\leavevmode\special{html:<a href="#1">}}%
 \providecommand\@@endlink[0]{\special{html:</a>}}%
}{%
 \providecommand\@@startlink[1]{%
  \leavevmode
  \pdfstartlink
   attr{/Border[0 0 1 ]/H/I/C[0 1 1]}%
   user{/Subtype/Link/A<</Type/Action/S/URI/URI(#1)>>}%
  \relax
 }%
 \providecommand\@@endlink[0]{\pdfendlink}%
}%
\providecommand \url  [0]{\begingroup\@sanitize \@url }%
\providecommand \@url [1]{\endgroup\@href {#1}{\urlprefix}}%
\providecommand \urlprefix [0]{URL }%
\providecommand \Eprint[0]{\href }%
\@ifxundefined \urlstyle {%
  \providecommand \doi [1]{doi:\discretionary{}{}{}#1}%
}{%
  \providecommand \doi [0]{doi:\discretionary{}{}{}\begingroup
  \urlstyle{rm}\Url }%
}%
\providecommand \doibase [0]{http://dx.doi.org/}%
\providecommand \Doi[1]{\href{\doibase#1}}%
\providecommand \bibAnnote [3]{%
  \BibitemShut{#1}%
  \begin{quotation}\noindent
    \textsc{Key:}\ #2\\\textsc{Annotation:}\ #3%
  \end{quotation}%
}%
\providecommand \bibAnnoteFile [2]{%
  \IfFileExists{#2}{\bibAnnote {#1} {#2} {\input{#2}}}{}%
}%
\providecommand \typeout [0]{\immediate \write \m@ne }%
\providecommand \selectlanguage [0]{\@gobble}%
\providecommand \bibinfo [0]{\@secondoftwo}%
\providecommand \bibfield [0]{\@secondoftwo}%
\providecommand \translation [1]{[#1]}%
\providecommand \BibitemOpen[0]{}%
\providecommand \bibitemStop [0]{}%
\providecommand \bibitemNoStop [0]{.\EOS\space}%
\providecommand \EOS [0]{\spacefactor3000\relax}%
\providecommand \BibitemShut [1]{\csname bibitem#1\endcsname}%
\bibitem{car_ice_92}%
  \BibitemOpen
  \bibfield{author}{%
  \bibinfo {author} {\bibfnamefont{C.}~\bibnamefont{Lee}}, \bibinfo {author}
  {\bibfnamefont{D.}~\bibnamefont{Vanderbilt}}, \bibinfo {author}
  {\bibfnamefont{K.}~\bibnamefont{Laasonen}}, \bibinfo {author}
  {\bibfnamefont{R.}~\bibnamefont{Car}},\ and\ \bibinfo {author}
  {\bibfnamefont{M.}~\bibnamefont{Parrinello}},\ }%
  \bibfield{journal}{%
  \bibinfo {journal} {Phys. Rev. Lett.}\ }%
  \textbf{\bibinfo {volume} {69}},\ \bibinfo {pages} {462} (\bibinfo {year}
  {1992})%
  \bibAnnoteFile{NoStop}{car_ice_92}%
\bibitem{klein_05}%
  \BibitemOpen
  \bibfield{author}{%
  \bibinfo {author} {\bibfnamefont{S.~J.}\ \bibnamefont{Singer}}, \bibinfo
  {author} {\bibfnamefont{J.-L.}\ \bibnamefont{Kuo}}, \bibinfo {author}
  {\bibfnamefont{T.~K.}\ \bibnamefont{Hirsch}}, \bibinfo {author}
  {\bibfnamefont{C.}~\bibnamefont{Knight}}, \bibinfo {author}
  {\bibfnamefont{L.}~\bibnamefont{Ojam{\"a}e}},\ and\ \bibinfo {author}
  {\bibfnamefont{M.~L.}\ \bibnamefont{Klein}},\ }%
  \bibfield{journal}{%
  \bibinfo {journal} {Phys. Rev. Lett.}\ }%
  \textbf{\bibinfo {volume} {94}},\ \bibinfo {pages} {135701} (\bibinfo {year}
  {2005})%
  \bibAnnoteFile{NoStop}{klein_05}%
\bibitem{slater_jacs_06}%
  \BibitemOpen
  \bibfield{author}{%
  \bibinfo {author} {\bibfnamefont{G.~A.}\ \bibnamefont{Tribello}}, \bibinfo
  {author} {\bibfnamefont{B.}~\bibnamefont{Slater}},\ and\ \bibinfo {author}
  {\bibfnamefont{C.~G.}\ \bibnamefont{Salzmann}},\ }%
  \bibfield{journal}{%
  \bibinfo {journal} {J. Am. Chem. Soc.}\ }%
  \textbf{\bibinfo {volume} {128}},\ \bibinfo {pages} {12594} (\bibinfo {year}
  {2006})%
  \bibAnnoteFile{NoStop}{slater_jacs_06}%
\bibitem{feibelman_08}%
  \BibitemOpen
  \bibfield{author}{%
  \bibinfo {author} {\bibfnamefont{P.~J.}\ \bibnamefont{Feibelman}},\ }%
  \bibfield{journal}{%
  \bibinfo {journal} {Phys. Chem. Chem. Phys.}\ }%
  \textbf{\bibinfo {volume} {10}},\ \bibinfo {pages} {4688} (\bibinfo {year}
  {2008})%
  \bibAnnoteFile{NoStop}{feibelman_08}%
\bibitem{hermann_prl_08}%
  \BibitemOpen
  \bibfield{author}{%
  \bibinfo {author} {\bibfnamefont{A.}~\bibnamefont{Hermann}}\ and\ \bibinfo
  {author} {\bibfnamefont{P.}~\bibnamefont{Schwerdtfeger}},\ }%
  \bibfield{journal}{%
  \bibinfo {journal} {Phys. Rev. Lett.}\ }%
  \textbf{\bibinfo {volume} {101}},\ \bibinfo {pages} {183005} (\bibinfo {year}
  {2008})%
  \bibAnnoteFile{NoStop}{hermann_prl_08}%
\bibitem{hamada_10}%
  \BibitemOpen
  \bibfield{author}{%
  \bibinfo {author} {\bibfnamefont{I.}~\bibnamefont{Hamada}},\ }%
  \bibfield{journal}{%
  \bibinfo {journal} {J. Chem. Phys.}\ }%
  \textbf{\bibinfo {volume} {133}},\ \bibinfo {pages} {214503} (\bibinfo {year}
  {2010})%
  \bibAnnoteFile{NoStop}{hamada_10}%
\bibitem{santra_hydrogen_2011}%
  \BibitemOpen
  \bibfield{author}{%
  \bibinfo {author} {\bibfnamefont{B.}~\bibnamefont{Santra}}, \bibinfo {author}
  {\bibfnamefont{J.}~\bibnamefont{Klime\v{s}}}, \bibinfo {author}
  {\bibfnamefont{D.}~\bibnamefont{Alf\`e}}, \bibinfo {author}
  {\bibfnamefont{A.}~\bibnamefont{Tkatchenko}}, \bibinfo {author}
  {\bibfnamefont{B.}~\bibnamefont{Slater}}, \bibinfo {author}
  {\bibfnamefont{A.}~\bibnamefont{Michaelides}}, \bibinfo {author}
  {\bibfnamefont{R.}~\bibnamefont{Car}},\ and\ \bibinfo {author}
  {\bibfnamefont{M.}~\bibnamefont{Scheffler}},\ }%
  \bibfield{journal}{%
  \bibinfo {journal} {Phys. Rev. Lett.}\ }%
  \textbf{\bibinfo {volume} {107}},\ \bibinfo {pages} {185701} (\bibinfo {year}
  {2011})%
  \bibAnnoteFile{NoStop}{santra_hydrogen_2011}%
\bibitem{murray_prl_2012}%
  \BibitemOpen
  \bibfield{author}{%
  \bibinfo {author} {\bibfnamefont{E.~D.}\ \bibnamefont{Murray}}\ and\ \bibinfo
  {author} {\bibfnamefont{G.}~\bibnamefont{Galli}},\ }%
  \bibfield{journal}{%
  \bibinfo {journal} {Phys. Rev. Lett.}\ }%
  \textbf{\bibinfo {volume} {108}},\ \bibinfo {pages} {105502} (\bibinfo {year}
  {2012})%
  \bibAnnoteFile{NoStop}{murray_prl_2012}%
\bibitem{pamuk_prl_2012}%
  \BibitemOpen
  \bibfield{author}{%
  \bibinfo {author} {\bibfnamefont{B.}~\bibnamefont{Pamuk}}, \bibinfo {author}
  {\bibfnamefont{J.~M.}\ \bibnamefont{Soler}}, \bibinfo {author}
  {\bibfnamefont{R.}~\bibnamefont{Ram\'irez}}, \bibinfo {author}
  {\bibfnamefont{C.~P.}\ \bibnamefont{Herrero}}, \bibinfo {author}
  {\bibfnamefont{P.}~\bibnamefont{Stephens}}, \bibinfo {author}
  {\bibfnamefont{P.~B.}\ \bibnamefont{Allen}},\ and\ \bibinfo {author}
  {\bibfnamefont{M.-V.}\ \bibnamefont{Fern\'andez-Serra}},\ }%
  \bibfield{journal}{%
  \bibinfo {journal} {Phys. Rev. Lett.}\ }%
  \textbf{\bibinfo {volume} {108}},\ \bibinfo {pages} {193003} (\bibinfo {year}
  {2012})%
  \bibAnnoteFile{NoStop}{pamuk_prl_2012}%
\bibitem{laasonen1993}%
  \BibitemOpen
  \bibfield{author}{%
  \bibinfo {author} {\bibfnamefont{K.}~\bibnamefont{Laasonen}}, \bibinfo
  {author} {\bibfnamefont{M.}~\bibnamefont{Sprik}}, \bibinfo {author}
  {\bibfnamefont{M.}~\bibnamefont{Parrinello}},\ and\ \bibinfo {author}
  {\bibfnamefont{R.}~\bibnamefont{Car}},\ }%
  \bibfield{journal}{%
  \bibinfo {journal} {J. Chem. Phys.}\ }%
  \textbf{\bibinfo {volume} {99}},\ \bibinfo {pages} {9080} (\bibinfo {year}
  {1993})%
  \bibAnnoteFile{NoStop}{laasonen1993}%
\bibitem{sprik_jcp_96}%
  \BibitemOpen
  \bibfield{author}{%
  \bibinfo {author} {\bibfnamefont{M.}~\bibnamefont{Sprik}}, \bibinfo {author}
  {\bibfnamefont{J.}~\bibnamefont{Hutter}},\ and\ \bibinfo {author}
  {\bibfnamefont{M.}~\bibnamefont{Parrinello}},\ }%
  \bibfield{journal}{%
  \bibinfo {journal} {J. Chem. Phys.}\ }%
  \textbf{\bibinfo {volume} {105}},\ \bibinfo {pages} {1142} (\bibinfo {year}
  {1996})%
  \bibAnnoteFile{NoStop}{sprik_jcp_96}%
\bibitem{silvestrelli_jcp_99}%
  \BibitemOpen
  \bibfield{author}{%
  \bibinfo {author} {\bibfnamefont{P.~L.}\ \bibnamefont{Silvestrelli}}\ and\
  \bibinfo {author} {\bibfnamefont{M.}~\bibnamefont{Parrinello}},\ }%
  \bibfield{journal}{%
  \bibinfo {journal} {J. Chem. Phys.}\ }%
  \textbf{\bibinfo {volume} {111}},\ \bibinfo {pages} {3572} (\bibinfo {year}
  {1999})%
  \bibAnnoteFile{NoStop}{silvestrelli_jcp_99}%
\bibitem{parrinello_03}%
  \BibitemOpen
  \bibfield{author}{%
  \bibinfo {author} {\bibfnamefont{B.}~\bibnamefont{Chen}}, \bibinfo {author}
  {\bibfnamefont{I.}~\bibnamefont{Ivanov}}, \bibinfo {author}
  {\bibfnamefont{M.~L.}\ \bibnamefont{Klein}},\ and\ \bibinfo {author}
  {\bibfnamefont{M.}~\bibnamefont{Parrinello}},\ }%
  \bibfield{journal}{%
  \bibinfo {journal} {Phys. Rev. Lett.}\ }%
  \textbf{\bibinfo {volume} {91}},\ \bibinfo {pages} {215503} (\bibinfo {year}
  {2003})%
  \bibAnnoteFile{NoStop}{parrinello_03}%
\bibitem{asthagiri_03}%
  \BibitemOpen
  \bibfield{author}{%
  \bibinfo {author} {\bibfnamefont{D.}~\bibnamefont{Asthagiri}}, \bibinfo
  {author} {\bibfnamefont{L.~R.}\ \bibnamefont{Pratt}},\ and\ \bibinfo {author}
  {\bibfnamefont{J.~D.}\ \bibnamefont{Kress}},\ }%
  \bibfield{journal}{%
  \bibinfo {journal} {Phys. Rev. E}\ }%
  \textbf{\bibinfo {volume} {68}},\ \bibinfo {pages} {041505} (\bibinfo {year}
  {2003})%
  \bibAnnoteFile{NoStop}{asthagiri_03}%
\bibitem{grossman_water_1}%
  \BibitemOpen
  \bibfield{author}{%
  \bibinfo {author} {\bibfnamefont{J.~C.}\ \bibnamefont{Grossman}}, \bibinfo
  {author} {\bibfnamefont{E.}~\bibnamefont{Schwegler}}, \bibinfo {author}
  {\bibfnamefont{E.~W.}\ \bibnamefont{Draeger}}, \bibinfo {author}
  {\bibfnamefont{F.}~\bibnamefont{Gygi}},\ and\ \bibinfo {author}
  {\bibfnamefont{G.}~\bibnamefont{Galli}},\ }%
  \bibfield{journal}{%
  \bibinfo {journal} {J. Chem. Phys.}\ }%
  \textbf{\bibinfo {volume} {120}},\ \bibinfo {pages} {300} (\bibinfo {year}
  {2004})%
  \bibAnnoteFile{NoStop}{grossman_water_1}%
\bibitem{grossman_water_2}%
  \BibitemOpen
  \bibfield{author}{%
  \bibinfo {author} {\bibfnamefont{E.}~\bibnamefont{Schwegler}}, \bibinfo
  {author} {\bibfnamefont{J.~C.}\ \bibnamefont{Grossman}}, \bibinfo {author}
  {\bibfnamefont{F.}~\bibnamefont{Gygi}},\ and\ \bibinfo {author}
  {\bibfnamefont{G.}~\bibnamefont{Galli}},\ }%
  \bibfield{journal}{%
  \bibinfo {journal} {J. Chem. Phys.}\ }%
  \textbf{\bibinfo {volume} {121}},\ \bibinfo {pages} {5400} (\bibinfo {year}
  {2004})%
  \bibAnnoteFile{NoStop}{grossman_water_2}%
\bibitem{artacho2004}%
  \BibitemOpen
  \bibfield{author}{%
  \bibinfo {author} {\bibfnamefont{M.~V.}\ \bibnamefont{Fern\'{a}ndez-Serra}}\
  and\ \bibinfo {author} {\bibfnamefont{E.}~\bibnamefont{Artacho}},\ }%
  \bibfield{journal}{%
  \bibinfo {journal} {J. Chem. Phys.}\ }%
  \textbf{\bibinfo {volume} {121}},\ \bibinfo {pages} {11136} (\bibinfo {year}
  {2004})%
  \bibAnnoteFile{NoStop}{artacho2004}%
\bibitem{Kuo_04}%
  \BibitemOpen
  \bibfield{author}{%
  \bibinfo {author} {\bibfnamefont{I.-F.~W.}\ \bibnamefont{Kuo}}, \bibinfo
  {author} {\bibfnamefont{C.~J.}\ \bibnamefont{Mundy}}, \bibinfo {author}
  {\bibfnamefont{M.~J.}\ \bibnamefont{McGrath}}, \bibinfo {author}
  {\bibfnamefont{J.~I.}\ \bibnamefont{Siepmann}}, \bibinfo {author}
  {\bibfnamefont{J.}~\bibnamefont{VandeVondele}}, \bibinfo {author}
  {\bibfnamefont{M.}~\bibnamefont{Sprik}}, \bibinfo {author}
  {\bibfnamefont{J.}~\bibnamefont{Hutter}}, \bibinfo {author}
  {\bibfnamefont{B.}~\bibnamefont{Chen}}, \bibinfo {author}
  {\bibfnamefont{M.~L.}\ \bibnamefont{Klein}}, \bibinfo {author}
  {\bibfnamefont{F.}~\bibnamefont{Mohamed}}, \bibinfo {author}
  {\bibfnamefont{M.}~\bibnamefont{Krack}},\ and\ \bibinfo {author}
  {\bibfnamefont{M.}~\bibnamefont{Parrinello}},\ }%
  \bibfield{journal}{%
  \bibinfo {journal} {J. Phys. Chem. B}\ }%
  \textbf{\bibinfo {volume} {108}},\ \bibinfo {pages} {12990} (\bibinfo {year}
  {2004})%
  \bibAnnoteFile{NoStop}{Kuo_04}%
\bibitem{Mcgrath_jpcA_06}%
  \BibitemOpen
  \bibfield{author}{%
  \bibinfo {author} {\bibfnamefont{M.~J.}\ \bibnamefont{McGrath}}, \bibinfo
  {author} {\bibfnamefont{J.~I.}\ \bibnamefont{Siepmann}}, \bibinfo {author}
  {\bibfnamefont{I.-F.~W.}\ \bibnamefont{Kuo}}, \bibinfo {author}
  {\bibfnamefont{C.~J.}\ \bibnamefont{Mundy}}, \bibinfo {author}
  {\bibfnamefont{J.}~\bibnamefont{VandeVondele}}, \bibinfo {author}
  {\bibfnamefont{J.}~\bibnamefont{Hutter}}, \bibinfo {author}
  {\bibfnamefont{F.}~\bibnamefont{Mohamed}},\ and\ \bibinfo {author}
  {\bibfnamefont{M.}~\bibnamefont{Krack}},\ }%
  \bibfield{journal}{%
  \bibinfo {journal} {J. Phys. Chem. A}\ }%
  \textbf{\bibinfo {volume} {110}},\ \bibinfo {pages} {640} (\bibinfo {year}
  {2006})%
  \bibAnnoteFile{NoStop}{Mcgrath_jpcA_06}%
\bibitem{Todorova_06}%
  \BibitemOpen
  \bibfield{author}{%
  \bibinfo {author} {\bibfnamefont{T.}~\bibnamefont{Todorova}}, \bibinfo
  {author} {\bibfnamefont{A.~P.}\ \bibnamefont{Seitsonen}}, \bibinfo {author}
  {\bibfnamefont{J.}~\bibnamefont{Hutter}}, \bibinfo {author}
  {\bibfnamefont{I.-F.~W.}\ \bibnamefont{Kuo}},\ and\ \bibinfo {author}
  {\bibfnamefont{C.~J.}\ \bibnamefont{Mundy}},\ }%
  \bibfield{journal}{%
  \bibinfo {journal} {J. Phys. Chem. B}\ }%
  \textbf{\bibinfo {volume} {110}},\ \bibinfo {pages} {3685} (\bibinfo {year}
  {2006})%
  \bibAnnoteFile{NoStop}{Todorova_06}%
\bibitem{tuckerman_jcp_06}%
  \BibitemOpen
  \bibfield{author}{%
  \bibinfo {author} {\bibfnamefont{H.-S.}\ \bibnamefont{Lee}}\ and\ \bibinfo
  {author} {\bibfnamefont{M.~E.}\ \bibnamefont{Tuckerman}},\ }%
  \bibfield{journal}{%
  \bibinfo {journal} {J. Chem. Phys.}\ }%
  \textbf{\bibinfo {volume} {125}},\ \bibinfo {pages} {154507} (\bibinfo {year}
  {2006})%
  \bibAnnoteFile{NoStop}{tuckerman_jcp_06}%
\bibitem{car_08}%
  \BibitemOpen
  \bibfield{author}{%
  \bibinfo {author} {\bibfnamefont{J.~A.}\ \bibnamefont{Morrone}}\ and\
  \bibinfo {author} {\bibfnamefont{R.}~\bibnamefont{Car}},\ }%
  \bibfield{journal}{%
  \bibinfo {journal} {Phys. Rev. Lett.}\ }%
  \textbf{\bibinfo {volume} {101}},\ \bibinfo {pages} {017801} (\bibinfo {year}
  {2008})%
  \bibAnnoteFile{NoStop}{car_08}%
\bibitem{parrinello_09}%
  \BibitemOpen
  \bibfield{author}{%
  \bibinfo {author} {\bibfnamefont{T.~D.}\ \bibnamefont{K\"{u}hne}}, \bibinfo
  {author} {\bibfnamefont{M.}~\bibnamefont{Krack}},\ and\ \bibinfo {author}
  {\bibfnamefont{M.}~\bibnamefont{Parrinello}},\ }%
  \bibfield{journal}{%
  \bibinfo {journal} {J. Chem. Theory Comput.}\ }%
  \textbf{\bibinfo {volume} {5}},\ \bibinfo {pages} {235} (\bibinfo {year}
  {2009})%
  \bibAnnoteFile{NoStop}{parrinello_09}%
\bibitem{xantheas_JCP_09}%
  \BibitemOpen
  \bibfield{author}{%
  \bibinfo {author} {\bibfnamefont{S.}~\bibnamefont{Yoo}}, \bibinfo {author}
  {\bibfnamefont{X.~C.}\ \bibnamefont{Zeng}},\ and\ \bibinfo {author}
  {\bibfnamefont{S.~S.}\ \bibnamefont{Xantheas}},\ }%
  \bibfield{journal}{%
  \bibinfo {journal} {J. Chem. Phys.}\ }%
  \textbf{\bibinfo {volume} {130}},\ \bibinfo {pages} {221102} (\bibinfo {year}
  {2009})%
  \bibAnnoteFile{NoStop}{xantheas_JCP_09}%
\bibitem{lin_jpcB_09}%
  \BibitemOpen
  \bibfield{author}{%
  \bibinfo {author} {\bibfnamefont{I.-C.}\ \bibnamefont{Lin}}, \bibinfo
  {author} {\bibfnamefont{A.~P.}\ \bibnamefont{Seitsonen}}, \bibinfo {author}
  {\bibfnamefont{M.~D.}\ \bibnamefont{Coutinho-Neto}}, \bibinfo {author}
  {\bibfnamefont{I.}~\bibnamefont{Tavernelli}},\ and\ \bibinfo {author}
  {\bibfnamefont{U.}~\bibnamefont{Rothlisberger}},\ }%
  \bibfield{journal}{%
  \bibinfo {journal} {J. Phys. Chem. B}\ }%
  \textbf{\bibinfo {volume} {113}},\ \bibinfo {pages} {1127} (\bibinfo {year}
  {2009})%
  \bibAnnoteFile{NoStop}{lin_jpcB_09}%
\bibitem{schmidt_jpcB_09}%
  \BibitemOpen
  \bibfield{author}{%
  \bibinfo {author} {\bibfnamefont{J.}~\bibnamefont{Schmidt}}, \bibinfo
  {author} {\bibfnamefont{J.}~\bibnamefont{VandeVondele}}, \bibinfo {author}
  {\bibfnamefont{I.-F.~W.}\ \bibnamefont{Kuo}}, \bibinfo {author}
  {\bibfnamefont{D.}~\bibnamefont{Sebastiani}}, \bibinfo {author}
  {\bibfnamefont{J.~I.}\ \bibnamefont{Siepmann}}, \bibinfo {author}
  {\bibfnamefont{J.}~\bibnamefont{Hutter}},\ and\ \bibinfo {author}
  {\bibfnamefont{C.~J.}\ \bibnamefont{Mundy}},\ }%
  \bibfield{journal}{%
  \bibinfo {journal} {J. Phys. Chem. B}\ }%
  \textbf{\bibinfo {volume} {113}},\ \bibinfo {pages} {11959} (\bibinfo {year}
  {2009})%
  \bibAnnoteFile{NoStop}{schmidt_jpcB_09}%
\bibitem{zhang_jctc_2011_2}%
  \BibitemOpen
  \bibfield{author}{%
  \bibinfo {author} {\bibfnamefont{C.}~\bibnamefont{Zhang}}, \bibinfo {author}
  {\bibfnamefont{D.}~\bibnamefont{Donadio}}, \bibinfo {author}
  {\bibfnamefont{F.}~\bibnamefont{Gygi}},\ and\ \bibinfo {author}
  {\bibfnamefont{G.}~\bibnamefont{Galli}},\ }%
  \bibfield{journal}{%
  \bibinfo {journal} {J. Chem. Theory Comput.}\ }%
  \textbf{\bibinfo {volume} {7}},\ \bibinfo {pages} {1443} (\bibinfo {year}
  {2011})%
  \bibAnnoteFile{NoStop}{zhang_jctc_2011_2}%
\bibitem{ojo_cpl_2011}%
  \BibitemOpen
  \bibfield{author}{%
  \bibinfo {author} {\bibfnamefont{O.}~\bibnamefont{Akin-Ojo}}\ and\ \bibinfo
  {author} {\bibfnamefont{F.}~\bibnamefont{Wang}},\ }%
  \bibfield{journal}{%
  \bibinfo {journal} {Chem. Phys. Lett.}\ }%
  \textbf{\bibinfo {volume} {513}},\ \bibinfo {pages} {59} (\bibinfo {year}
  {2011})%
  \bibAnnoteFile{NoStop}{ojo_cpl_2011}%
\bibitem{wang_jcp_2011}%
  \BibitemOpen
  \bibfield{author}{%
  \bibinfo {author} {\bibfnamefont{J.}~\bibnamefont{Wang}}, \bibinfo {author}
  {\bibfnamefont{G.}~\bibnamefont{Rom\'{a}n-P\'{e}rez}}, \bibinfo {author}
  {\bibfnamefont{J.~M.}\ \bibnamefont{Soler}}, \bibinfo {author}
  {\bibfnamefont{E.}~\bibnamefont{Artacho}},\ and\ \bibinfo {author}
  {\bibfnamefont{M.-V.}\ \bibnamefont{Fern\'{a}ndez-Serra}},\ }%
  \bibfield{journal}{%
  \bibinfo {journal} {J. Chem. Phys.}\ }%
  \textbf{\bibinfo {volume} {134}},\ \bibinfo {pages} {024516} (\bibinfo {year}
  {2011})%
  \bibAnnoteFile{NoStop}{wang_jcp_2011}%
\bibitem{yoo_jcp_11}%
  \BibitemOpen
  \bibfield{author}{%
  \bibinfo {author} {\bibfnamefont{S.}~\bibnamefont{Yoo}}\ and\ \bibinfo
  {author} {\bibfnamefont{S.~S.}\ \bibnamefont{Xantheas}},\ }%
  \bibfield{journal}{%
  \bibinfo {journal} {J. Chem. Phys.}\ }%
  \textbf{\bibinfo {volume} {134}},\ \bibinfo {pages} {121105} (\bibinfo {year}
  {2012})%
  \bibAnnoteFile{NoStop}{yoo_jcp_11}%
\bibitem{zhang_jctc_2011}%
  \BibitemOpen
  \bibfield{author}{%
  \bibinfo {author} {\bibfnamefont{C.}~\bibnamefont{Zhang}}, \bibinfo {author}
  {\bibfnamefont{J.}~\bibnamefont{Wu}}, \bibinfo {author}
  {\bibfnamefont{G.}~\bibnamefont{Galli}},\ and\ \bibinfo {author}
  {\bibfnamefont{F.}~\bibnamefont{Gygi}},\ }%
  \bibfield{journal}{%
  \bibinfo {journal} {J. Chem. Theory Comput.}\ }%
  \textbf{\bibinfo {volume} {7}},\ \bibinfo {pages} {3054} (\bibinfo {year}
  {2011})%
  \bibAnnoteFile{NoStop}{zhang_jctc_2011}%
\bibitem{kim_jordan_1994}%
  \BibitemOpen
  \bibfield{author}{%
  \bibinfo {author} {\bibfnamefont{K.}~\bibnamefont{Kim}}\ and\ \bibinfo
  {author} {\bibfnamefont{K.~D.}\ \bibnamefont{Jordan}},\ }%
  \bibfield{journal}{%
  \bibinfo {journal} {J. Phys. Chem.}\ }%
  \textbf{\bibinfo {volume} {98}},\ \bibinfo {pages} {10089} (\bibinfo {year}
  {1994})%
  \bibAnnoteFile{NoStop}{kim_jordan_1994}%
\bibitem{xantheas_1995}%
  \BibitemOpen
  \bibfield{author}{%
  \bibinfo {author} {\bibfnamefont{S.~S.}\ \bibnamefont{Xantheas}},\ }%
  \bibfield{journal}{%
  \bibinfo {journal} {J. Chem. Phys.}\ }%
  \textbf{\bibinfo {volume} {102}},\ \bibinfo {pages} {4505} (\bibinfo {year}
  {1995})%
  \bibAnnoteFile{NoStop}{xantheas_1995}%
\bibitem{xantheas_JCP_02}%
  \BibitemOpen
  \bibfield{author}{%
  \bibinfo {author} {\bibfnamefont{S.~S.}\ \bibnamefont{Xantheas}}, \bibinfo
  {author} {\bibfnamefont{C.~J.}\ \bibnamefont{Burnham}},\ and\ \bibinfo
  {author} {\bibfnamefont{R.~J.}\ \bibnamefont{Harrison}},\ }%
  \bibfield{journal}{%
  \bibinfo {journal} {J. Chem. Phys.}\ }%
  \textbf{\bibinfo {volume} {116}},\ \bibinfo {pages} {116} (\bibinfo {year}
  {2002})%
  \bibAnnoteFile{NoStop}{xantheas_JCP_02}%
\bibitem{santra_jcp_2007}%
  \BibitemOpen
  \bibfield{author}{%
  \bibinfo {author} {\bibfnamefont{B.}~\bibnamefont{Santra}}, \bibinfo {author}
  {\bibfnamefont{A.}~\bibnamefont{Michaelides}},\ and\ \bibinfo {author}
  {\bibfnamefont{M.}~\bibnamefont{Scheffler}},\ }%
  \bibfield{journal}{%
  \bibinfo {journal} {J. Chem. Phys.}\ }%
  \textbf{\bibinfo {volume} {127}},\ \bibinfo {pages} {184104} (\bibinfo {year}
  {2007})%
  \bibAnnoteFile{NoStop}{santra_jcp_2007}%
\bibitem{santra_jcp_2009}%
  \BibitemOpen
  \bibfield{author}{%
  \bibinfo {author} {\bibfnamefont{B.}~\bibnamefont{Santra}}, \bibinfo {author}
  {\bibfnamefont{A.}~\bibnamefont{Michaelides}},\ and\ \bibinfo {author}
  {\bibfnamefont{M.}~\bibnamefont{Scheffler}},\ }%
  \bibfield{journal}{%
  \bibinfo {journal} {J. Chem. Phys.}\ }%
  \textbf{\bibinfo {volume} {131}},\ \bibinfo {pages} {124509} (\bibinfo {year}
  {2009})%
  \bibAnnoteFile{NoStop}{santra_jcp_2009}%
\bibitem{truhlar_hexamer_2008}%
  \BibitemOpen
  \bibfield{author}{%
  \bibinfo {author} {\bibfnamefont{E.~E.}\ \bibnamefont{Dahlke}}, \bibinfo
  {author} {\bibfnamefont{R.~M.}\ \bibnamefont{Olson}}, \bibinfo {author}
  {\bibfnamefont{H.~R.}\ \bibnamefont{Leverentz}},\ and\ \bibinfo {author}
  {\bibfnamefont{D.~G.}\ \bibnamefont{Truhlar}},\ }%
  \bibfield{journal}{%
  \bibinfo {journal} {J. Phys. Chem. A}\ }%
  \textbf{\bibinfo {volume} {112}},\ \bibinfo {pages} {3976} (\bibinfo {year}
  {2008})%
  \bibAnnoteFile{NoStop}{truhlar_hexamer_2008}%
\bibitem{shields_kirschner_2008}%
  \BibitemOpen
  \bibfield{author}{%
  \bibinfo {author} {\bibfnamefont{G.~C.}\ \bibnamefont{Shields}}\ and\
  \bibinfo {author} {\bibfnamefont{K.~N.}\ \bibnamefont{Kirschner}},\ }%
  \bibfield{journal}{%
  \bibinfo {journal} {Synthesis and Reactivity in Inorganic, Metal-Organic, and
  Nano-Metal Chemistry}\ }%
  \textbf{\bibinfo {volume} {38}},\ \bibinfo {pages} {32} (\bibinfo {year}
  {2008})%
  \bibAnnoteFile{NoStop}{shields_kirschner_2008}%
\bibitem{hammond_09}%
  \BibitemOpen
  \bibfield{author}{%
  \bibinfo {author} {\bibfnamefont{J.~R.}\ \bibnamefont{Hammond}}, \bibinfo
  {author} {\bibfnamefont{N.}~\bibnamefont{Govind}}, \bibinfo {author}
  {\bibfnamefont{K.}~\bibnamefont{Kowalski}}, \bibinfo {author}
  {\bibfnamefont{J.}~\bibnamefont{Autschbach}},\ and\ \bibinfo {author}
  {\bibfnamefont{S.~S.}\ \bibnamefont{Xantheas}},\ }%
  \bibfield{journal}{%
  \bibinfo {journal} {J. Chem. Phys.}\ }%
  \textbf{\bibinfo {volume} {131}},\ \bibinfo {pages} {214103} (\bibinfo {year}
  {2009})%
  \bibAnnoteFile{NoStop}{hammond_09}%
\bibitem{jordan_10}%
  \BibitemOpen
  \bibfield{author}{%
  \bibinfo {author} {\bibfnamefont{F.-F.}\ \bibnamefont{Wang}}, \bibinfo
  {author} {\bibfnamefont{G.}~\bibnamefont{Jenness}}, \bibinfo {author}
  {\bibfnamefont{W.~A.}\ \bibnamefont{Al-Saidi}},\ and\ \bibinfo {author}
  {\bibfnamefont{K.~D.}\ \bibnamefont{Jordan}},\ }%
  \bibfield{journal}{%
  \bibinfo {journal} {J. Chem. Phys.}\ }%
  \textbf{\bibinfo {volume} {132}},\ \bibinfo {pages} {134303} (\bibinfo {year}
  {2010})%
  \bibAnnoteFile{NoStop}{jordan_10}%
\bibitem{shields_jpca_2010}%
  \BibitemOpen
  \bibfield{author}{%
  \bibinfo {author} {\bibfnamefont{R.~M.}\ \bibnamefont{Shields}}, \bibinfo
  {author} {\bibfnamefont{B.}~\bibnamefont{Temelso}}, \bibinfo {author}
  {\bibfnamefont{K.~A.}\ \bibnamefont{Archer}}, \bibinfo {author}
  {\bibfnamefont{T.~E.}\ \bibnamefont{Morrell}},\ and\ \bibinfo {author}
  {\bibfnamefont{G.~C.}\ \bibnamefont{Shields}},\ }%
  \bibfield{journal}{%
  \bibinfo {journal} {J. Phys. Chem. A}\ }%
  \textbf{\bibinfo {volume} {114}},\ \bibinfo {pages} {11725} (\bibinfo {year}
  {2010})%
  \bibAnnoteFile{NoStop}{shields_jpca_2010}%
\bibitem{bygrave_jcp_12}%
  \BibitemOpen
  \bibfield{author}{%
  \bibinfo {author} {\bibfnamefont{P.~J.}\ \bibnamefont{Bygrave}}, \bibinfo
  {author} {\bibfnamefont{N.~L.}\ \bibnamefont{Allan}},\ and\ \bibinfo {author}
  {\bibfnamefont{F.~R.}\ \bibnamefont{Manby}},\ }%
  \bibfield{journal}{%
  \bibinfo {journal} {J. Chem. Phys.}\ }%
  \textbf{\bibinfo {volume} {137}},\ \bibinfo {pages} {164102} (\bibinfo {year}
  {2012})%
  \bibAnnoteFile{NoStop}{bygrave_jcp_12}%
\bibitem{gillan_jcp_12}%
  \BibitemOpen
  \bibfield{author}{%
  \bibinfo {author} {\bibfnamefont{M.~J.}\ \bibnamefont{Gillan}}, \bibinfo
  {author} {\bibfnamefont{F.~R.}\ \bibnamefont{Manby}}, \bibinfo {author}
  {\bibfnamefont{M.~D.}\ \bibnamefont{Towler}},\ and\ \bibinfo {author}
  {\bibfnamefont{D.}~\bibnamefont{Alf\`e}},\ }%
  \bibfield{journal}{%
  \bibinfo {journal} {J. Chem. Phys.}\ }%
  \textbf{\bibinfo {volume} {136}},\ \bibinfo {pages} {244105} (\bibinfo {year}
  {2012})%
  \bibAnnoteFile{NoStop}{gillan_jcp_12}%
\bibitem{gillan_jcp_13a}%
  \BibitemOpen
  \bibfield{author}{%
  \bibinfo {author} {\bibfnamefont{M.~J.}\ \bibnamefont{Gillan}}, \bibinfo
  {author} {\bibfnamefont{D.}~\bibnamefont{Alf\`e}}, \bibinfo {author}
  {\bibfnamefont{P.~J.}\ \bibnamefont{Bygrave}}, \bibinfo {author}
  {\bibfnamefont{C.~R.}\ \bibnamefont{Taylor}},\ and\ \bibinfo {author}
  {\bibfnamefont{F.~R.}\ \bibnamefont{Manby}},\ }%
  \bibfield{journal}{%
  \bibinfo {journal} {J. Chem. Phys.}\ }%
  \textbf{\bibinfo {volume} {139}},\ \bibinfo {pages} {114101} (\bibinfo {year}
  {2013})%
  \bibAnnoteFile{NoStop}{gillan_jcp_13a}%
\bibitem{delben2013jpcl}%
  \BibitemOpen
  \bibfield{author}{%
  \bibinfo {author} {\bibfnamefont{M.~D.}\ \bibnamefont{Ben}}, \bibinfo
  {author} {\bibfnamefont{M.}~\bibnamefont{Sch\"{o}nherr}}, \bibinfo {author}
  {\bibfnamefont{J.}~\bibnamefont{Hutter}},\ and\ \bibinfo {author}
  {\bibfnamefont{J.}~\bibnamefont{VandeVondele}},\ }%
  \bibfield{journal}{%
  \bibinfo {journal} {J. Phys. Chem. Lett.}\ }%
  \textbf{\bibinfo {volume} {4}},\ \bibinfo {pages} {3753} (\bibinfo {year}
  {2013})%
  \bibAnnoteFile{NoStop}{delben2013jpcl}%
\bibitem{delben2013jctc}%
  \BibitemOpen
  \bibfield{author}{%
  \bibinfo {author} {\bibfnamefont{M.~D.}\ \bibnamefont{Ben}}, \bibinfo
  {author} {\bibfnamefont{J.}~\bibnamefont{Hutter}},\ and\ \bibinfo {author}
  {\bibfnamefont{J.}~\bibnamefont{VandeVondele}},\ }%
  \bibfield{journal}{%
  \bibinfo {journal} {J. Chem. Theo. Comput.}\ }%
  \textbf{\bibinfo {volume} {9}},\ \bibinfo {pages} {2654} (\bibinfo {year}
  {2013})%
  \bibAnnoteFile{NoStop}{delben2013jctc}%
\bibitem{manby2010chap}%
  \BibitemOpen
  \bibfield{author}{%
  \bibinfo {author} {\bibfnamefont{D.~P.}\ \bibnamefont{O'Neill}}, \bibinfo
  {author} {\bibfnamefont{N.~L.}\ \bibnamefont{Allan}},\ and\ \bibinfo {author}
  {\bibfnamefont{F.~R.}\ \bibnamefont{Manby}},\ }%
  in\ \emph{\bibinfo {booktitle} {Accurate Condensed-Phase Quantum
  Chemistry}},\ \bibinfo {editor} {edited by\ \bibinfo {editor}
  {\bibfnamefont{F.}~\bibnamefont{Manby}}}\ (\bibinfo {publisher} {CRC Press,
  Taylor \& Francis Group, Boca Raton},\ \bibinfo {year} {2010})%
  \bibAnnoteFile{NoStop}{manby2010chap}%
\bibitem{dion2004}%
  \BibitemOpen
  \bibfield{author}{%
  \bibinfo {author} {\bibfnamefont{M.}~\bibnamefont{Dion}}, \bibinfo {author}
  {\bibfnamefont{H.}~\bibnamefont{Rydberg}}, \bibinfo {author}
  {\bibfnamefont{E.}~\bibnamefont{Schr\"oder}}, \bibinfo {author}
  {\bibfnamefont{D.~C.}\ \bibnamefont{Langreth}},\ and\ \bibinfo {author}
  {\bibfnamefont{B.~I.}\ \bibnamefont{Lundqvist}},\ }%
  \bibfield{journal}{%
  \bibinfo {journal} {Phys. Rev. Lett.}\ }%
  \textbf{\bibinfo {volume} {92}},\ \bibinfo {pages} {246401} (\bibinfo {year}
  {2004})%
  \bibAnnoteFile{NoStop}{dion2004}%
\bibitem{langreth2005}%
  \BibitemOpen
  \bibfield{author}{%
  \bibinfo {author} {\bibfnamefont{D.~C.}\ \bibnamefont{Langreth}}, \bibinfo
  {author} {\bibfnamefont{M.}~\bibnamefont{Dion}}, \bibinfo {author}
  {\bibfnamefont{H.}~\bibnamefont{Rydberg}}, \bibinfo {author}
  {\bibfnamefont{E.}~\bibnamefont{Schroder}}, \bibinfo {author}
  {\bibfnamefont{P.}~\bibnamefont{Hyldgaard}},\ and\ \bibinfo {author}
  {\bibfnamefont{B.~I.}\ \bibnamefont{Lundqvist}},\ }%
  \bibfield{journal}{%
  \bibinfo {journal} {Int. J. Quant. Chem.}\ }%
  \textbf{\bibinfo {volume} {{101}}},\ \bibinfo {pages} {{599}} (\bibinfo
  {year} {{2005}})%
  \bibAnnoteFile{NoStop}{langreth2005}%
\bibitem{kelkkanen2009}%
  \BibitemOpen
  \bibfield{author}{%
  \bibinfo {author} {\bibfnamefont{A.~K.}\ \bibnamefont{Kelkkanen}}, \bibinfo
  {author} {\bibfnamefont{B.~I.}\ \bibnamefont{Lundqvist}},\ and\ \bibinfo
  {author} {\bibfnamefont{J.~K.}\ \bibnamefont{N{\o}rskov}},\ }%
  \bibfield{journal}{%
  \bibinfo {journal} {J. Chem. Phys.}\ }%
  \textbf{\bibinfo {volume} {131}},\ \bibinfo {pages} {046102} (\bibinfo {year}
  {2009})%
  \bibAnnoteFile{NoStop}{kelkkanen2009}%
\bibitem{grimme2006d2}%
  \BibitemOpen
  \bibfield{author}{%
  \bibinfo {author} {\bibfnamefont{S.}~\bibnamefont{Grimme}},\ }%
  \bibfield{journal}{%
  \bibinfo {journal} {J. Comput. Chem.}\ }%
  \textbf{\bibinfo {volume} {27}},\ \bibinfo {pages} {1787} (\bibinfo {year}
  {2006})%
  \bibAnnoteFile{NoStop}{grimme2006d2}%
\bibitem{tkatchenko2009}%
  \BibitemOpen
  \bibfield{author}{%
  \bibinfo {author} {\bibfnamefont{A.}~\bibnamefont{Tkatchenko}}\ and\ \bibinfo
  {author} {\bibfnamefont{M.}~\bibnamefont{Scheffler}},\ }%
  \bibfield{journal}{%
  \bibinfo {journal} {Phys. Rev. Lett.}\ }%
  \textbf{\bibinfo {volume} {102}},\ \bibinfo {pages} {073005} (\bibinfo {year}
  {2009})%
  \bibAnnoteFile{NoStop}{tkatchenko2009}%
\bibitem{santra2008}%
  \BibitemOpen
  \bibfield{author}{%
  \bibinfo {author} {\bibfnamefont{B.}~\bibnamefont{Santra}}, \bibinfo {author}
  {\bibfnamefont{A.}~\bibnamefont{Michaelides}}, \bibinfo {author}
  {\bibfnamefont{M.}~\bibnamefont{Fuchs}}, \bibinfo {author}
  {\bibfnamefont{A.}~\bibnamefont{Tkatchenko}}, \bibinfo {author}
  {\bibfnamefont{C.}~\bibnamefont{Filippi}},\ and\ \bibinfo {author}
  {\bibfnamefont{M.}~\bibnamefont{Scheffler}},\ }%
  \bibfield{journal}{%
  \bibinfo {journal} {J. Chem. Phys.}\ }%
  \textbf{\bibinfo {volume} {129}},\ \bibinfo {pages} {194111} (\bibinfo {year}
  {2008})%
  \bibAnnoteFile{NoStop}{santra2008}%
\bibitem{klimes2012}%
  \BibitemOpen
  \bibfield{author}{%
  \bibinfo {author} {\bibfnamefont{J.}~\bibnamefont{Klime\v{s}}}\ and\ \bibinfo
  {author} {\bibfnamefont{A.}~\bibnamefont{Michaelides}},\ }%
  \bibfield{journal}{%
  \bibinfo {journal} {J. Chem. Phys.}\ }%
  \textbf{\bibinfo {volume} {137}},\ \bibinfo {pages} {120901} (\bibinfo {year}
  {2012})%
  \bibAnnoteFile{NoStop}{klimes2012}%
\bibitem{carrasco2011}%
  \BibitemOpen
  \bibfield{author}{%
  \bibinfo {author} {\bibfnamefont{J.}~\bibnamefont{Carrasco}}, \bibinfo
  {author} {\bibfnamefont{B.}~\bibnamefont{Santra}}, \bibinfo {author}
  {\bibfnamefont{J.}~\bibnamefont{Klime\v{s}}},\ and\ \bibinfo {author}
  {\bibfnamefont{A.}~\bibnamefont{Michaelides}},\ }%
  \bibfield{journal}{%
  \bibinfo {journal} {Phys. Rev. Lett.}\ }%
  \textbf{\bibinfo {volume} {106}},\ \bibinfo {pages} {026101} (\bibinfo {year}
  {2011})%
  \bibAnnoteFile{NoStop}{carrasco2011}%
\bibitem{carrasco2012nmat}%
  \BibitemOpen
  \bibfield{author}{%
  \bibinfo {author} {\bibfnamefont{J.}~\bibnamefont{Carrasco}}, \bibinfo
  {author} {\bibfnamefont{A.}~\bibnamefont{Hodgson}},\ and\ \bibinfo {author}
  {\bibfnamefont{A.}~\bibnamefont{Michaelides}},\ }%
  \bibfield{journal}{%
  \bibinfo {journal} {Nature Mater.}\ }%
  \textbf{\bibinfo {volume} {11}},\ \bibinfo {pages} {667} (\bibinfo {year}
  {2012})%
  \bibAnnoteFile{NoStop}{carrasco2012nmat}%
\bibitem{carrasco2013}%
  \BibitemOpen
  \bibfield{author}{%
  \bibinfo {author} {\bibfnamefont{J.}~\bibnamefont{Carrasco}}, \bibinfo
  {author} {\bibfnamefont{J.}~\bibnamefont{Klime\v{s}}},\ and\ \bibinfo
  {author} {\bibfnamefont{A.}~\bibnamefont{Michaelides}},\ }%
  \bibfield{journal}{%
  \bibinfo {journal} {J. Chem. Phys.}\ }%
  \textbf{\bibinfo {volume} {138}},\ \bibinfo {pages} {024708} (\bibinfo {year}
  {2013})%
  \bibAnnoteFile{NoStop}{carrasco2013}%
\bibitem{harl2008}%
  \BibitemOpen
  \bibfield{author}{%
  \bibinfo {author} {\bibfnamefont{J.}~\bibnamefont{Harl}}\ and\ \bibinfo
  {author} {\bibfnamefont{G.}~\bibnamefont{Kresse}},\ }%
  \bibfield{journal}{%
  \bibinfo {journal} {Phys. Rev. B}\ }%
  \textbf{\bibinfo {volume} {77}},\ \bibinfo {pages} {045136} (\bibinfo {year}
  {2008})%
  \bibAnnoteFile{NoStop}{harl2008}%
\bibitem{lebegue2010}%
  \BibitemOpen
  \bibfield{author}{%
  \bibinfo {author} {\bibfnamefont{S.}~\bibnamefont{Leb\`egue}}, \bibinfo
  {author} {\bibfnamefont{J.}~\bibnamefont{Harl}}, \bibinfo {author}
  {\bibfnamefont{T.}~\bibnamefont{Gould}}, \bibinfo {author}
  {\bibfnamefont{J.~G.}\ \bibnamefont{\'Angy\'an}}, \bibinfo {author}
  {\bibfnamefont{G.}~\bibnamefont{Kresse}},\ and\ \bibinfo {author}
  {\bibfnamefont{J.~F.}\ \bibnamefont{Dobson}},\ }%
  \bibfield{journal}{%
  \bibinfo {journal} {Phys. Rev. Lett.}\ }%
  \textbf{\bibinfo {volume} {105}},\ \bibinfo {pages} {196401} (\bibinfo {year}
  {2010})%
  \bibAnnoteFile{NoStop}{lebegue2010}%
\bibitem{ren2011}%
  \BibitemOpen
  \bibfield{author}{%
  \bibinfo {author} {\bibfnamefont{X.}~\bibnamefont{Ren}}, \bibinfo {author}
  {\bibfnamefont{A.}~\bibnamefont{Tkatchenko}}, \bibinfo {author}
  {\bibfnamefont{P.}~\bibnamefont{Rinke}},\ and\ \bibinfo {author}
  {\bibfnamefont{M.}~\bibnamefont{Scheffler}},\ }%
  \bibfield{journal}{%
  \bibinfo {journal} {Phys. Rev. Lett.}\ }%
  \textbf{\bibinfo {volume} {106}},\ \bibinfo {pages} {153003} (\bibinfo {year}
  {2011})%
  \bibAnnoteFile{NoStop}{ren2011}%
\bibitem{harl2009}%
  \BibitemOpen
  \bibfield{author}{%
  \bibinfo {author} {\bibfnamefont{J.}~\bibnamefont{Harl}}\ and\ \bibinfo
  {author} {\bibfnamefont{G.}~\bibnamefont{Kresse}},\ }%
  \bibfield{journal}{%
  \bibinfo {journal} {Phys. Rev. Lett.}\ }%
  \textbf{\bibinfo {volume} {103}},\ \bibinfo {pages} {056401} (\bibinfo {year}
  {2010})%
  \bibAnnoteFile{NoStop}{harl2009}%
\bibitem{harl2010}%
  \BibitemOpen
  \bibfield{author}{%
  \bibinfo {author} {\bibfnamefont{J.}~\bibnamefont{Harl}}, \bibinfo {author}
  {\bibfnamefont{L.}~\bibnamefont{Schimka}},\ and\ \bibinfo {author}
  {\bibfnamefont{G.}~\bibnamefont{Kresse}},\ }%
  \bibfield{journal}{%
  \bibinfo {journal} {Phys. Rev. B}\ }%
  \textbf{\bibinfo {volume} {81}},\ \bibinfo {pages} {115126} (\bibinfo {year}
  {2010})%
  \bibAnnoteFile{NoStop}{harl2010}%
\bibitem{schimka2013}%
  \BibitemOpen
  \bibfield{author}{%
  \bibinfo {author} {\bibfnamefont{L.}~\bibnamefont{Schimka}}, \bibinfo
  {author} {\bibfnamefont{R.}~\bibnamefont{Gaudoin}}, \bibinfo {author}
  {\bibfnamefont{J.}~\bibnamefont{Klime\v{s}}}, \bibinfo {author}
  {\bibfnamefont{M.}~\bibnamefont{Marsman}},\ and\ \bibinfo {author}
  {\bibfnamefont{G.}~\bibnamefont{Kresse}},\ }%
  \bibfield{journal}{%
  \bibinfo {journal} {Phys. Rev. B}\ }%
  \textbf{\bibinfo {volume} {87}},\ \bibinfo {pages} {214102} (\bibinfo {year}
  {2013})%
  \bibAnnoteFile{NoStop}{schimka2013}%
\bibitem{yan2013}%
  \BibitemOpen
  \bibfield{author}{%
  \bibinfo {author} {\bibfnamefont{J.}~\bibnamefont{Yan}}, \bibinfo {author}
  {\bibfnamefont{J.~S.}\ \bibnamefont{Hummelsh{\o}j}},\ and\ \bibinfo {author}
  {\bibfnamefont{J.~K.}\ \bibnamefont{N{\o}rskov}},\ }%
  \bibfield{journal}{%
  \bibinfo {journal} {Phys. Rev. B}\ }%
  \textbf{\bibinfo {volume} {87}},\ \bibinfo {pages} {075207} (\bibinfo {year}
  {2013})%
  \bibAnnoteFile{NoStop}{yan2013}%
\bibitem{blochl1994}%
  \BibitemOpen
  \bibfield{author}{%
  \bibinfo {author} {\bibfnamefont{P.~E.}\ \bibnamefont{Bl\"ochl}},\ }%
  \bibfield{journal}{%
  \bibinfo {journal} {Phys. Rev. B}\ }%
  \textbf{\bibinfo {volume} {50}},\ \bibinfo {pages} {17953} (\bibinfo {year}
  {1994})%
  \bibAnnoteFile{NoStop}{blochl1994}%
\bibitem{kresse1999}%
  \BibitemOpen
  \bibfield{author}{%
  \bibinfo {author} {\bibfnamefont{G.}~\bibnamefont{Kresse}}\ and\ \bibinfo
  {author} {\bibfnamefont{J.}~\bibnamefont{Joubert}},\ }%
  \bibfield{journal}{%
  \bibinfo {journal} {Phys. Rev. B}\ }%
  \textbf{\bibinfo {volume} {59}},\ \bibinfo {pages} {1758} (\bibinfo {year}
  {1999})%
  \bibAnnoteFile{NoStop}{kresse1999}%
\bibitem{perdew1996}%
  \BibitemOpen
  \bibfield{author}{%
  \bibinfo {author} {\bibfnamefont{J.~P.}\ \bibnamefont{Perdew}}, \bibinfo
  {author} {\bibfnamefont{K.}~\bibnamefont{Burke}},\ and\ \bibinfo {author}
  {\bibfnamefont{M.}~\bibnamefont{Ernzerhof}},\ }%
  \bibfield{journal}{%
  \bibinfo {journal} {Phys. Rev. Lett.}\ }%
  \textbf{\bibinfo {volume} {77}},\ \bibinfo {pages} {3865} (\bibinfo {year}
  {1996}),\ \bibinfo {note} {$ibid$, {\bf 78}, 1396 (1997)}%
  \bibAnnoteFile{NoStop}{perdew1996}%
\bibitem{paier2012}%
  \BibitemOpen
  \bibfield{author}{%
  \bibinfo {author} {\bibfnamefont{J.}~\bibnamefont{Paier}}, \bibinfo {author}
  {\bibfnamefont{X.}~\bibnamefont{Ren}}, \bibinfo {author}
  {\bibfnamefont{P.}~\bibnamefont{Rinke}}, \bibinfo {author}
  {\bibfnamefont{G.~E.}\ \bibnamefont{Scuseria}}, \bibinfo {author}
  {\bibfnamefont{A.}~\bibnamefont{Gr\"{u}neis}}, \bibinfo {author}
  {\bibfnamefont{G.}~\bibnamefont{Kresse}},\ and\ \bibinfo {author}
  {\bibfnamefont{M.}~\bibnamefont{Scheffler}},\ }%
  \bibfield{journal}{%
  \bibinfo {journal} {New J. Phys.}\ }%
  \textbf{\bibinfo {volume} {14}},\ \bibinfo {pages} {043002} (\bibinfo {year}
  {2012})%
  \bibAnnoteFile{NoStop}{paier2012}%
\bibitem{ren2012}%
  \BibitemOpen
  \bibfield{author}{%
  \bibinfo {author} {\bibfnamefont{X.}~\bibnamefont{Ren}}, \bibinfo {author}
  {\bibfnamefont{P.}~\bibnamefont{Rinke}}, \bibinfo {author}
  {\bibfnamefont{V.}~\bibnamefont{Blum}}, \bibinfo {author}
  {\bibfnamefont{J.}~\bibnamefont{Wieferink}}, \bibinfo {author}
  {\bibfnamefont{A.}~\bibnamefont{Tkatchenko}}, \bibinfo {author}
  {\bibfnamefont{A.}~\bibnamefont{Sanfilippo}}, \bibinfo {author}
  {\bibfnamefont{K.}~\bibnamefont{Reuter}},\ and\ \bibinfo {author}
  {\bibfnamefont{M.}~\bibnamefont{Scheffler}},\ }%
  \bibfield{journal}{%
  \bibinfo {journal} {New J. Phys.}\ }%
  \textbf{\bibinfo {volume} {14}},\ \bibinfo {pages} {053020} (\bibinfo {year}
  {2012})%
  \bibAnnoteFile{NoStop}{ren2012}%
\bibitem{ren_renormalized_2013}%
  \BibitemOpen
  \bibfield{author}{%
  \bibinfo {author} {\bibfnamefont{X.}~\bibnamefont{Ren}}, \bibinfo {author}
  {\bibfnamefont{P.}~\bibnamefont{Rinke}}, \bibinfo {author}
  {\bibfnamefont{G.~E.}\ \bibnamefont{Scuseria}},\ and\ \bibinfo {author}
  {\bibfnamefont{M.}~\bibnamefont{Scheffler}},\ }%
  \bibfield{journal}{%
  \bibinfo {journal} {Phys. Rev. B}\ }%
  \textbf{\bibinfo {volume} {88}},\ \bibinfo {pages} {035120} (\bibinfo {year}
  {2013})%
  \bibAnnoteFile{NoStop}{ren_renormalized_2013}%
\bibitem{leadbetter_equilibrium_1985}%
  \BibitemOpen
  \bibfield{author}{%
  \bibinfo {author} {\bibfnamefont{A.~J.}\ \bibnamefont{Leadbetter}}, \bibinfo
  {author} {\bibfnamefont{R.~C.}\ \bibnamefont{Ward}}, \bibinfo {author}
  {\bibfnamefont{J.~W.}\ \bibnamefont{Clark}}, \bibinfo {author}
  {\bibfnamefont{P.~A.}\ \bibnamefont{Tucker}}, \bibinfo {author}
  {\bibfnamefont{T.}~\bibnamefont{Matsuo}},\ and\ \bibinfo {author}
  {\bibfnamefont{H.}~\bibnamefont{Suga}},\ }%
  \bibfield{journal}{%
  \bibinfo {journal} {J. Chem. Phys.}\ }%
  \textbf{\bibinfo {volume} {82}},\ \bibinfo {pages} {424} (\bibinfo {year}
  {1985})%
  \bibAnnoteFile{NoStop}{leadbetter_equilibrium_1985}%
\bibitem{howe_determination_1989}%
  \BibitemOpen
  \bibfield{author}{%
  \bibinfo {author} {\bibfnamefont{R.}~\bibnamefont{Howe}}\ and\ \bibinfo
  {author} {\bibfnamefont{R.~W.}\ \bibnamefont{Whitworth}},\ }%
  \bibfield{journal}{%
  \bibinfo {journal} {J. Chem. Phys.}\ }%
  \textbf{\bibinfo {volume} {90}},\ \bibinfo {pages} {4450} (\bibinfo {year}
  {1989})%
  \bibAnnoteFile{NoStop}{howe_determination_1989}%
\bibitem{hamann_h_2o_1997}%
  \BibitemOpen
  \bibfield{author}{%
  \bibinfo {author} {\bibfnamefont{D.~R.}\ \bibnamefont{Hamann}},\ }%
  \bibfield{journal}{%
  \Doi{10.1103/PhysRevB.55.R10157}{\bibinfo {journal} {Phys. Rev. B}}\ }%
  \textbf{\bibinfo {volume} {55}},\ \bibinfo {pages} {R10157} (\bibinfo {year}
  {1997})%
  \bibAnnoteFile{NoStop}{hamann_h_2o_1997}%
\bibitem{santra_accuracy_2013}%
  \BibitemOpen
  \bibfield{author}{%
  \bibinfo {author} {\bibfnamefont{B.}~\bibnamefont{Santra}}, \bibinfo {author}
  {\bibfnamefont{J.}~\bibnamefont{Klime\v{s}}}, \bibinfo {author}
  {\bibfnamefont{A.}~\bibnamefont{Tkatchenko}}, \bibinfo {author}
  {\bibfnamefont{D.}~\bibnamefont{Alf\`{e}}}, \bibinfo {author}
  {\bibfnamefont{B.}~\bibnamefont{Slater}}, \bibinfo {author}
  {\bibfnamefont{A.}~\bibnamefont{Michaelides}}, \bibinfo {author}
  {\bibfnamefont{R.}~\bibnamefont{Car}},\ and\ \bibinfo {author}
  {\bibfnamefont{M.}~\bibnamefont{Scheffler}},\ }%
  \bibfield{journal}{%
  \bibinfo {journal} {J. Chem. Phys.}\ }%
  \textbf{\bibinfo {volume} {139}},\ \bibinfo {pages} {154702} (\bibinfo {year}
  {2013})%
  \bibAnnoteFile{NoStop}{santra_accuracy_2013}%
\bibitem{davidson_proposed_1984}%
  \BibitemOpen
  \bibfield{author}{%
  \bibinfo {author} {\bibfnamefont{E.~R.}\ \bibnamefont{Davidson}}\ and\
  \bibinfo {author} {\bibfnamefont{K.}~\bibnamefont{Morokuma}},\ }%
  \bibfield{journal}{%
  \Doi{10.1063/1.448101}{\bibinfo {journal} {J. Chem. Phys.}}\ }%
  \textbf{\bibinfo {volume} {81}},\ \bibinfo {pages} {3741} (\bibinfo {year}
  {1984})%
  \bibAnnoteFile{NoStop}{davidson_proposed_1984}%
\bibitem{hirsch_quantum-chemical_2004}%
  \BibitemOpen
  \bibfield{author}{%
  \bibinfo {author} {\bibfnamefont{T.~K.}\ \bibnamefont{Hirsch}}\ and\ \bibinfo
  {author} {\bibfnamefont{L.}~\bibnamefont{Ojam\"{a}e}},\ }%
  \bibfield{journal}{%
  \Doi{10.1021/jp048434u}{\bibinfo {journal} {J. Phys. Chem. B}}\ }%
  \textbf{\bibinfo {volume} {108}},\ \bibinfo {pages} {15856} (\bibinfo {year}
  {2004})%
  \bibAnnoteFile{NoStop}{hirsch_quantum-chemical_2004}%
\bibitem{Profeta}%
  \BibitemOpen
  \bibfield{author}{%
  \bibinfo {author} {\bibfnamefont{G.}~\bibnamefont{Profeta}}\ and\ \bibinfo
  {author} {\bibfnamefont{S.}~\bibnamefont{Scandolo}},\ }%
  \bibfield{journal}{%
  \Doi{10.1103/PhysRevB.84.024103}{\bibinfo {journal} {Phys. Rev. B}}\ }%
  \textbf{\bibinfo {volume} {84}},\ \bibinfo {pages} {024103} (\bibinfo {year}
  {2011})%
  \bibAnnoteFile{NoStop}{Profeta}%
\bibitem{murray_formation_2005}%
  \BibitemOpen
  \bibfield{author}{%
  \bibinfo {author} {\bibfnamefont{B.~J.}\ \bibnamefont{Murray}}, \bibinfo
  {author} {\bibfnamefont{D.~A.}\ \bibnamefont{Knopf}},\ and\ \bibinfo {author}
  {\bibfnamefont{A.~K.}\ \bibnamefont{Bertram}},\ }%
  \bibfield{journal}{%
  \Doi{10.1038/nature03403}{\bibinfo {journal} {Nature}}\ }%
  \textbf{\bibinfo {volume} {434}},\ \bibinfo {pages} {202} (\bibinfo {year}
  {2005})%
  \bibAnnoteFile{NoStop}{murray_formation_2005}%
\bibitem{raza_proton_2011}%
  \BibitemOpen
  \bibfield{author}{%
  \bibinfo {author} {\bibfnamefont{Z.}~\bibnamefont{Raza}}, \bibinfo {author}
  {\bibfnamefont{D.}~\bibnamefont{Alf\'{e}}}, \bibinfo {author}
  {\bibfnamefont{C.~G.}\ \bibnamefont{Salzmann}}, \bibinfo {author}
  {\bibfnamefont{J.}~\bibnamefont{Klime\v{s}}}, \bibinfo {author}
  {\bibfnamefont{A.}~\bibnamefont{Michaelides}},\ and\ \bibinfo {author}
  {\bibfnamefont{B.}~\bibnamefont{Slater}},\ }%
  \bibfield{journal}{%
  \Doi{10.1039/c1cp22506e}{\bibinfo {journal} {Phys. Chem. Chem. Phys.}}\ }%
  \textbf{\bibinfo {volume} {13}},\ \bibinfo {pages} {19788} (\bibinfo {year}
  {2011})%
  \bibAnnoteFile{NoStop}{raza_proton_2011}%
\bibitem{geiger2013}%
  \BibitemOpen
  \bibfield{author}{%
  \bibinfo {author} {\bibfnamefont{P.}~\bibnamefont{Geiger}}, \bibinfo {author}
  {\bibfnamefont{C.}~\bibnamefont{Dellago}}, \bibinfo {author}
  {\bibfnamefont{M.}~\bibnamefont{Macher}}, \bibinfo {author}
  {\bibfnamefont{C.}~\bibnamefont{Franchini}}, \bibinfo {author}
  {\bibfnamefont{G.}~\bibnamefont{Kresse}}, \bibinfo {author}
  {\bibfnamefont{J.}~\bibnamefont{Bernard}}, \bibinfo {author}
  {\bibfnamefont{J.~N.}\ \bibnamefont{Stern}},\ and\ \bibinfo {author}
  {\bibfnamefont{T.}~\bibnamefont{Loerting}}\ }%
  \bibinfo {note} {{\it submitted}}%
  \bibAnnoteFile{NoStop}{geiger2013}%
\bibitem{whalley_ice_1968}%
  \BibitemOpen
  \bibfield{author}{%
  \bibinfo {author} {\bibfnamefont{E.}~\bibnamefont{Whalley}}, \bibinfo
  {author} {\bibfnamefont{J.~B.~R.}\ \bibnamefont{Heath}},\ and\ \bibinfo
  {author} {\bibfnamefont{D.~W.}\ \bibnamefont{Davidson}},\ }%
  \bibfield{journal}{%
  \Doi{doi:10.1063/1.1669438}{\bibinfo {journal} {J. Chem. Phys.}}\ }%
  \textbf{\bibinfo {volume} {48}},\ \bibinfo {pages} {2362} (\bibinfo {year}
  {1968})%
  \bibAnnoteFile{NoStop}{whalley_ice_1968}%
\bibitem{kamb_ice._1964}%
  \BibitemOpen
  \bibfield{author}{%
  \bibinfo {author} {\bibfnamefont{B.}~\bibnamefont{Kamb}},\ }%
  \bibfield{journal}{%
  \Doi{10.1107/S0365110X64003553}{\bibinfo {journal} {Acta Crystallographica}}\
  }%
  \textbf{\bibinfo {volume} {17}},\ \bibinfo {pages} {1437} (\bibinfo {year}
  {1964})%
  \bibAnnoteFile{NoStop}{kamb_ice._1964}%
\bibitem{chaplin_water_2012}%
  \BibitemOpen
  \bibfield{author}{%
  \bibinfo {author} {\bibfnamefont{M.}~\bibnamefont{Chaplin}},\ }%
  \enquote{\bibinfo {title} {Water phase diagram},}\  (\bibinfo {year}
  {2012}),\ \bibinfo {note} {http://www.lsbu.ac.uk/water/phase.html, accessed
  2. 7. 2012}%
  \bibAnnoteFile{NoStop}{chaplin_water_2012}%
\bibitem{salzmann_preparation_2006}%
  \BibitemOpen
  \bibfield{author}{%
  \bibinfo {author} {\bibfnamefont{C.~G.}\ \bibnamefont{Salzmann}},\ }%
  \bibfield{journal}{%
  \Doi{10.1126/science.1123896}{\bibinfo {journal} {Science}}\ }%
  \textbf{\bibinfo {volume} {311}},\ \bibinfo {pages} {1758} (\bibinfo {year}
  {2006})%
  \bibAnnoteFile{NoStop}{salzmann_preparation_2006}%
\bibitem{salzmann_ice_2009}%
  \BibitemOpen
  \bibfield{author}{%
  \bibinfo {author} {\bibfnamefont{C.}~\bibnamefont{Salzmann}}, \bibinfo
  {author} {\bibfnamefont{P.}~\bibnamefont{Radaelli}}, \bibinfo {author}
  {\bibfnamefont{E.}~\bibnamefont{Mayer}},\ and\ \bibinfo {author}
  {\bibfnamefont{J.}~\bibnamefont{Finney}},\ }%
  \bibfield{journal}{%
  \bibinfo {journal} {Phys. Rev. Lett.}\ }%
  \textbf{\bibinfo {volume} {103}},\ \bibinfo {pages} {105701} (\bibinfo {year}
  {2009})%
  \bibAnnoteFile{NoStop}{salzmann_ice_2009}%
\bibitem{kuhs_structure_1984}%
  \BibitemOpen
  \bibfield{author}{%
  \bibinfo {author} {\bibfnamefont{W.~F.}\ \bibnamefont{Kuhs}}, \bibinfo
  {author} {\bibfnamefont{J.~L.}\ \bibnamefont{Finney}}, \bibinfo {author}
  {\bibfnamefont{C.}~\bibnamefont{Vettier}},\ and\ \bibinfo {author}
  {\bibfnamefont{D.~V.}\ \bibnamefont{Bliss}},\ }%
  \bibfield{journal}{%
  \Doi{doi:10.1063/1.448109}{\bibinfo {journal} {J. Chem. Phys.}}\ }%
  \textbf{\bibinfo {volume} {81}},\ \bibinfo {pages} {3612} (\bibinfo {year}
  {1984})%
  \bibAnnoteFile{NoStop}{kuhs_structure_1984}%
\bibitem{pruzan_stability_1993}%
  \BibitemOpen
  \bibfield{author}{%
  \bibinfo {author} {\bibfnamefont{P.}~\bibnamefont{Pruzan}}, \bibinfo {author}
  {\bibfnamefont{J.~C.}\ \bibnamefont{Chervin}},\ and\ \bibinfo {author}
  {\bibfnamefont{B.}~\bibnamefont{Canny}},\ }%
  \bibfield{journal}{%
  \Doi{10.1063/1.465467}{\bibinfo {journal} {J. Chem. Phys.}}\ }%
  \textbf{\bibinfo {volume} {99}},\ \bibinfo {pages} {9842} (\bibinfo {year}
  {1993})%
  \bibAnnoteFile{NoStop}{pruzan_stability_1993}%
\bibitem{jorgensen_structure_1984}%
  \BibitemOpen
  \bibfield{author}{%
  \bibinfo {author} {\bibfnamefont{J.~D.}\ \bibnamefont{Jorgensen}}, \bibinfo
  {author} {\bibfnamefont{R.~A.}\ \bibnamefont{Beyerlein}}, \bibinfo {author}
  {\bibfnamefont{N.}~\bibnamefont{Watanabe}},\ and\ \bibinfo {author}
  {\bibfnamefont{T.~G.}\ \bibnamefont{Worlton}},\ }%
  \bibfield{journal}{%
  \Doi{10.1063/1.448027}{\bibinfo {journal} {J. Chem. Phys.}}\ }%
  \textbf{\bibinfo {volume} {81}},\ \bibinfo {pages} {3211} (\bibinfo {year}
  {1984})%
  \bibAnnoteFile{NoStop}{jorgensen_structure_1984}%
\bibitem{yoshimura_high-pressure_2006}%
  \BibitemOpen
  \bibfield{author}{%
  \bibinfo {author} {\bibfnamefont{Y.}~\bibnamefont{Yoshimura}}, \bibinfo
  {author} {\bibfnamefont{S.~T.}\ \bibnamefont{Stewart}}, \bibinfo {author}
  {\bibfnamefont{M.}~\bibnamefont{Somayazulu}}, \bibinfo {author}
  {\bibfnamefont{H.-k.}\ \bibnamefont{Mao}},\ and\ \bibinfo {author}
  {\bibfnamefont{R.~J.}\ \bibnamefont{Hemley}},\ }%
  \bibfield{journal}{%
  \Doi{doi:10.1063/1.2140277}{\bibinfo {journal} {J. Chem. Phys.}}\ }%
  \textbf{\bibinfo {volume} {124}},\ \bibinfo {pages} {024502} (\bibinfo {year}
  {2006})%
  \bibAnnoteFile{NoStop}{yoshimura_high-pressure_2006}%
\bibitem{la_placa_nearly_1973}%
  \BibitemOpen
  \bibfield{author}{%
  \bibinfo {author} {\bibfnamefont{S.~J.}\ \bibnamefont{La~Placa}}, \bibinfo
  {author} {\bibfnamefont{W.~C.}\ \bibnamefont{Hamilton}}, \bibinfo {author}
  {\bibfnamefont{B.}~\bibnamefont{Kamb}},\ and\ \bibinfo {author}
  {\bibfnamefont{A.}~\bibnamefont{Prakash}},\ }%
  \bibfield{journal}{%
  \Doi{doi:10.1063/1.1679238}{\bibinfo {journal} {J. Chem. Phys.}}\ }%
  \textbf{\bibinfo {volume} {58}},\ \bibinfo {pages} {567} (\bibinfo {year}
  {1973})%
  \bibAnnoteFile{NoStop}{la_placa_nearly_1973}%
\bibitem{londono_neutron_1993}%
  \BibitemOpen
  \bibfield{author}{%
  \bibinfo {author} {\bibfnamefont{J.~D.}\ \bibnamefont{Londono}}, \bibinfo
  {author} {\bibfnamefont{W.~F.}\ \bibnamefont{Kuhs}},\ and\ \bibinfo {author}
  {\bibfnamefont{J.~L.}\ \bibnamefont{Finney}},\ }%
  \bibfield{journal}{%
  \bibinfo {journal} {J. Chem. Phys.}\ }%
  \textbf{\bibinfo {volume} {98}},\ \bibinfo {pages} {4878} (\bibinfo {year}
  {1993})%
  \bibAnnoteFile{NoStop}{londono_neutron_1993}%
\bibitem{whalley_energies_1984}%
  \BibitemOpen
  \bibfield{author}{%
  \bibinfo {author} {\bibfnamefont{E.}~\bibnamefont{Whalley}},\ }%
  \bibfield{journal}{%
  \Doi{doi:10.1063/1.448153}{\bibinfo {journal} {J. Chem. Phys.}}\ }%
  \textbf{\bibinfo {volume} {81}},\ \bibinfo {pages} {4087} (\bibinfo {year}
  {1984})%
  \bibAnnoteFile{NoStop}{whalley_energies_1984}%
\bibitem{Lekner1997}%
  \BibitemOpen
  \bibfield{author}{%
  \bibinfo {author} {\bibfnamefont{J.}~\bibnamefont{Lekner}},\ }%
  \bibfield{journal}{%
  \Doi{http://dx.doi.org/10.1016/S0921-4526(97)00430-4}{\bibinfo {journal}
  {Physica B: Condensed Matter}}\ }%
  \textbf{\bibinfo {volume} {240}},\ \bibinfo {pages} {263 } (\bibinfo {year}
  {1997})%
  \bibAnnoteFile{NoStop}{Lekner1997}%
\bibitem{adamo1999}%
  \BibitemOpen
  \bibfield{author}{%
  \bibinfo {author} {\bibfnamefont{C.}~\bibnamefont{Adamo}}\ and\ \bibinfo
  {author} {\bibfnamefont{V.}~\bibnamefont{Barone}},\ }%
  \bibfield{journal}{%
  \bibinfo {journal} {J. Chem. Phys.}\ }%
  \textbf{\bibinfo {volume} {110}},\ \bibinfo {pages} {6158} (\bibinfo {year}
  {1999})%
  \bibAnnoteFile{NoStop}{adamo1999}%
\bibitem{heyd2003hse}%
  \BibitemOpen
  \bibfield{author}{%
  \bibinfo {author} {\bibfnamefont{J.}~\bibnamefont{Heyd}}, \bibinfo {author}
  {\bibfnamefont{G.~E.}\ \bibnamefont{Scuseria}},\ and\ \bibinfo {author}
  {\bibfnamefont{M.}~\bibnamefont{Ernzerhof}},\ }%
  \bibfield{journal}{%
  \bibinfo {journal} {J. Chem. Phys.}\ }%
  \textbf{\bibinfo {volume} {{118}}},\ \bibinfo {pages} {{8207}} (\bibinfo
  {year} {{2003}})%
  \bibAnnoteFile{NoStop}{heyd2003hse}%
\bibitem{heyd2006hse}%
  \BibitemOpen
  \bibfield{author}{%
  \bibinfo {author} {\bibfnamefont{J.}~\bibnamefont{Heyd}}, \bibinfo {author}
  {\bibfnamefont{G.~E.}\ \bibnamefont{Scuseria}},\ and\ \bibinfo {author}
  {\bibfnamefont{M.}~\bibnamefont{Ernzerhof}},\ }%
  \bibfield{journal}{%
  \bibinfo {journal} {{J. Chem. Phys.}}\ }%
  \textbf{\bibinfo {volume} {{124}}},\ \bibinfo {pages} {{219906}} (\bibinfo
  {year} {{2006}})%
  \bibAnnoteFile{NoStop}{heyd2006hse}%
\bibitem{hobbs_ice_1974}%
  \BibitemOpen
  \bibfield{author}{%
  \bibinfo {author} {\bibfnamefont{P.~V.}\ \bibnamefont{Hobbs}},\ }%
  \emph{\bibinfo {title} {Ice physics}},\ Oxford classic texts in the physical
  sciences\ (\bibinfo {publisher} {Oxford University Press},\ \bibinfo
  {address} {New York},\ \bibinfo {year} {1974})\ ISBN \bibinfo {isbn}
  {9780199587711}%
  \bibAnnoteFile{NoStop}{hobbs_ice_1974}%
\bibitem{vega_radial_2005}%
  \BibitemOpen
  \bibfield{author}{%
  \bibinfo {author} {\bibfnamefont{C.}~\bibnamefont{Vega}}, \bibinfo {author}
  {\bibfnamefont{C.}~\bibnamefont{{McBride}}}, \bibinfo {author}
  {\bibfnamefont{E.}~\bibnamefont{Sanz}},\ and\ \bibinfo {author}
  {\bibfnamefont{J.~L.~F.}\ \bibnamefont{Abascal}},\ }%
  \bibfield{journal}{%
  \Doi{10.1039/b418934e}{\bibinfo {journal} {Phys. Chem. Chem. Phys.}}\ }%
  \textbf{\bibinfo {volume} {7}},\ \bibinfo {pages} {1450} (\bibinfo {year}
  {2005})%
  \bibAnnoteFile{NoStop}{vega_radial_2005}%
\bibitem{line_high_1996}%
  \BibitemOpen
  \bibfield{author}{%
  \bibinfo {author} {\bibfnamefont{C.~M.~B.}\ \bibnamefont{Line}}\ and\
  \bibinfo {author} {\bibfnamefont{R.~W.}\ \bibnamefont{Whitworth}},\ }%
  \bibfield{journal}{%
  \Doi{10.1063/1.471745}{\bibinfo {journal} {J. Chem. Phys.}}\ }%
  \textbf{\bibinfo {volume} {104}},\ \bibinfo {pages} {10008} (\bibinfo {year}
  {1996})%
  \bibAnnoteFile{NoStop}{line_high_1996}%
\bibitem{fortes_incompressibility_2005}%
  \BibitemOpen
  \bibfield{author}{%
  \bibinfo {author} {\bibfnamefont{A.~D.}\ \bibnamefont{Fortes}}, \bibinfo
  {author} {\bibfnamefont{I.~G.}\ \bibnamefont{Wood}}, \bibinfo {author}
  {\bibfnamefont{M.}~\bibnamefont{Alfredsson}}, \bibinfo {author}
  {\bibfnamefont{L.}~\bibnamefont{Vo\v{c}adlo}},\ and\ \bibinfo {author}
  {\bibfnamefont{K.~S.}\ \bibnamefont{Knight}},\ }%
  \bibfield{journal}{%
  \Doi{10.1107/S0021889805014226}{\bibinfo {journal} {Journal of Applied
  Crystallography}}\ }%
  \textbf{\bibinfo {volume} {38}},\ \bibinfo {pages} {612} (\bibinfo {year}
  {2005})%
  \bibAnnoteFile{NoStop}{fortes_incompressibility_2005}%
\bibitem{lobban_pt_2002}%
  \BibitemOpen
  \bibfield{author}{%
  \bibinfo {author} {\bibfnamefont{C.}~\bibnamefont{Lobban}}, \bibinfo {author}
  {\bibfnamefont{J.~L.}\ \bibnamefont{Finney}},\ and\ \bibinfo {author}
  {\bibfnamefont{W.~F.}\ \bibnamefont{Kuhs}},\ }%
  \bibfield{journal}{%
  \Doi{doi:10.1063/1.1495837}{\bibinfo {journal} {J. Chem. Phys.}}\ }%
  \textbf{\bibinfo {volume} {117}},\ \bibinfo {pages} {3928} (\bibinfo {year}
  {2002})%
  \bibAnnoteFile{NoStop}{lobban_pt_2002}%
\end{thebibliography}%

\end{document}